\documentclass[a4paper,11pt]{article}
\pdfoutput=1 

\usepackage{jcappub} 

\usepackage[T1]{fontenc} 
\usepackage{comment}

\newcommand{\dis}[1]{\begin{equation}\begin{split}#1\end{split}\end{equation}}

\newcommand{\bfrac}[2]{\left(\frac{#1}{#2} \right)  }
\newcommand{\eq}[1]{Eq.~(\ref{#1})}

\title{\boldmath A title with some math: $x=1$}

\author[a]{Ki-Young Choi,}
\author[b]{Jongkuk Kim,}
\author[a]{and Erdenebulgan Lkhagvadorj}

\affiliation[a]{Department of Physics and Institute of Basic Science, Sungkyunkwan University, 2066 Seobu-ro, Suwon-si, Gyeonggi-do, 16419, Korea}
\affiliation[b]{School of Physics, KIAS, 85 Hoegiro, Seoul 02455, Korea}

\emailAdd{kiyoungchoi@skku.edu}
\emailAdd{jkkim@kias.re.kr}
\emailAdd{bulgaa@skku.edu}

\abstract{
We consider the possibility that the weakly interacting massive particles produced from the evaporation of primordial black hole  can  explain both the relic density of dark matter and the baryon asymmetry of the Universe, through their annihilation which violate  B and CP-symmetry. We find that the primordial black hole with mass less than $10^7 {\rm g}$ is a good candidate as an source of TeV dark matter with the total annihilation cross section $\left\langle\sigma_a \upsilon\right\rangle \lesssim 10^{-7} \ {\rm GeV}^{-2}$ and the B-violating scattering cross section $\left\langle\sigma_B \upsilon\right\rangle \lesssim 2\times 10^{-9} \ {\rm GeV^{-2}}$. 
This large annihilation cross section of dark matter in this model would make it available to search them in the indirect search for dark matter such as gamma-ray or neutrino observations.
}

\hypersetup{
colorlinks = true,
linkcolor = blue,
citecolor = magenta
}

\begin{document}

\begin{flushright}
KIAS-P23044
\end{flushright}

\maketitle
\flushbottom
\section{Introduction}

There are many observational evidences that the dark matter (DM) constitutes approximately 27\% of the present Universe's energy density, which is reported as $\Omega_{\rm DM} h^2 = 0.120\pm 0.001$ at 68\% Confidence Level~\cite{Planck:2018vyg}.
The DM abundance is nearly five times the relic density of ordinary matter or baryons. The baryonic matter content, comprising 5\% of the present Universe's energy density, is highly asymmetric giving rise to the longstanding puzzle of the baryon asymmetry of the Universe (BAU). 
The observed BAU is quoted as the ratio of excess of baryons over anti-baryons to photon~\cite{Planck:2018vyg, PDG2020}
\begin{equation}
    \eta_{B} \equiv \frac{n_B - n_{\bar{B}}}{n_\gamma} \simeq 6.2 \times 10^{-10},
\end{equation}
or  the corresponding comoving baryon asymmetry is $Y_B \equiv (n_B - n_{\bar{B}})/s \simeq 8.7 \times 10^{-11}$, with the entropy density $s$. However, the Standard Model (SM) of particle physics fails to explain both the nature of DM and the criteria for generating BAU known as Sakharov's conditions~\cite{Sakharov_1991}. 
Many studies have been conducted to address both problems in beyond the Standard Model. One of the simplest and most natural models is WIMPy baryogenesis~\cite{Cui:2011ab, Bernal:2012gv, Bernal:2013bga} where Weakly Interacting Massive Particle (WIMP) dark matter freezes out to generate its own relic abundance and whose annihilation can play a crucial role in generating the baryon asymmetry during thermal freeze-out.  

Another interesting possibility in the early Universe is the formation of small-mass primordial black holes (PBH) and their evaporation~\cite{HAWKING1974, HAWKING1975}. The PBHs are formed through various mechanisms, such as the collapse of primordial density inhomogeneities originating from quantum fluctuations prior to inflation, or from cosmic strings, domain walls, or bubble collisions during a first order phase transition~\cite{Khlopov:2008qy, Carr_2010, Carr_2021, Carr:2021bzv, Villanueva-Domingo:2021spv}. Moreover, PBH evaporation can contribute to the production of DM and the generation of baryon asymmetry in various distinct scenarios~\cite{Fujita2014, Morrison:2018xla, Masina:2020xhk, Datta:2020bht, Bernal:2021kaj, JyotiDas:2021shi, Bernal:2022pue, Barman:2022gjo, Barman:2021ost,Bhaumik:2022pil, Gehrman:2022imk, ShamsEsHaghi:2022azq, Barman:2022pdo}.  In our study, we consider pre-existing PBHs that are produced by some of these processes in the earlier Universe.
Once PBHs are formed, they behave as cold matter and could have constituted dominant fraction of the energy density of Universe, indicating early matter-dominated epoch (EMDE).
We focus on their evaporation into WIMP DM, which may become the dominant component of DM and also their annihilation may generate the BAU. 

In this paper, we assume that all of PBHs consist of the type of Schwarzschild black holes with a single mass initially, independent of the spin and angular momentum. 
We consider PBHs with a mass below $10^9$ g, enabling them to decay prior Big Bang nucleosynthesis (BBN)~\cite{Carr_2010, Carr_2021, Papanikolaou:2020qtd}. 
Moreover, we are interested in the PBH which decay after thermal freeze-out of WIMP DM, allowing non-thermal production of DM to make a dominant contribution to the whole DM composition. 
This requires that the evaporation temperature of PBH is lower than the freeze-out temperature of WIMP, usually given by $T_{\rm FO}\simeq m_{\rm DM} / (20\sim25)$. 
Therefore, the non-thermal DM can naturally satisfy  the out-of-equilibrium, which is one of the Sakharov conditions.

Finally, if the annihilation of WIMP contains B-violation and C, CP-violation, it becomes possible to generate the baryon asymmetry during the annihilation of the non-thermal WIMPs.
To calculate the DM production and successive generation of the baryon asymmetry, we numerically solve a set of the Boltzmann equations that incorporate the PBH evaporation and DM annihilation using a publicly available python package\footnote{\url{https://github.com/earlyuniverse/ulysses}}~\cite{Granelli:2020pim} for this purpose. Futhermore, we also provide some semi-analytic solutions that are consistent with our numerical result.

Importantly, the re-annihilation of non-thermally generated WIMP can successfully explain both the correct DM relic abundance and the observed baryon asymmetry simultaneously~\cite{CHOI2018657}. In comparison to the thermal WIMPy baryogenesis, the non-thermally produced baryogenesis accumulates significantly during EMDE. 

The paper is organized as follows. In Sec.~\ref{section2}, we briefly introduce the PBH formation and evaporation properties, highlighting their behavior as Schwarzschild black holes. Additionally, some analytical estimation for the non-thermal WIMP coming from PBHs is discussed. In Sec.~\ref{section3}, we establish the Boltzmann equations that we solve with the presence of PBHs and DMs. We then proceed to describe
our results for WIMPy baryogenesis with the possible washout effects.
In Sec.~\ref{IDDM}, we briefly comment on the possible constraints and signals of our model in the indirect searches of DM.
Finally, we make our concluding remarks in Sec.~\ref{con}.

\section{Primordial Black Hole: Generation and Evaporation} \label{section2}

When a sufficiently large density perturbation reenters the Hubble horizon, it undergoes gravitational collapse and forms a black hole. PBHs have been formed after inflation during radiation-dominated epoch~\cite{Carr_2010, Carr_2021, Carr:2020xqk}. In the radiation-dominated Universe, the initial mass of black hole, $M_{\rm in}$, at the time of its formation can be related to the background energy density $\rho$ and the Hubble parameter $H$ at the corresponding temperature $T_{\rm in}$, which is given by
~\cite{Fujita2014, Masina:2020xhk, Gondolo2020}
\begin{equation}\label{eq:MassPBH}
    M_{\rm in} \equiv M_{\rm BH}(T_{\text{in}}) = \gamma \frac{4 \pi}{3} \frac{\rho(T_{\rm in})}{H^3(T_{\rm in})},
\end{equation}
   where $\gamma\simeq 0.2$ is a numerical correction factor, $\rho(T_{\rm in}) = 3 M_p^2 H^2(T_{\rm in})$ is total energy density with the reduced Planck mass $M_p = 2.4 \times 10^{18} \ \rm{GeV}\simeq 4.3 \times 10^{-6} \ \rm{g}$. Consequently, the temperature of the plasma $T_{\rm in}$ can be written in terms of $M_{\rm in}$ as
\begin{equation}\label{eq:initialTem}
    T_{\rm in} = \left(\frac{1440 \ \gamma^2}{g_*(T_{\rm in})}\right)^{1/4} \sqrt{\frac{M_p^3}{M_{\rm in}}},
\end{equation}
where $g_*(T_{\rm in})$ is the effective degrees of freedom of the energy density at the formation temperature.

Once formed, PBH can evaporate into particles by emitting Hawking radiation, if their masses are smaller than the instantaneous Hawking temperature of PBH, $T_{\text{BH}}$, given by~\cite{HAWKING1974, HAWKING1975} 
\begin{equation}\label{eq:Hawkingtemperature}
    T_\text{BH} = \frac{M_p^2}{M_\text{BH}}\simeq 10^7 \rm GeV \left(\frac{10^6 \ \rm g}{M_{\rm BH}}\right).
\end{equation}
In the case of Schwarzschild black hole, the emission rate of particles with momentum $p$  becomes~\cite{UKWATTA201690, Lunardini:2019zob, Perez-Gonzalez:2020vnz}:
\begin{equation}\label{eq:emission rate}
    \frac{d^2N_i}{dp \ dt} = \frac{g_i}{2 \pi^2} \frac{\sigma_{i}(M_\text{BH}, \mu_i,p)}{e^{E_i(p)/T_\text{BH}} \pm 1} \frac{p^3}{E_i(p)}, 
\end{equation}
where $g_i$ represents the number of degrees of freedom of $i$-th species, $E_i(p) = \sqrt{\mu_i^2 + p^2}$ is total energy of particles with mass $\mu_i$, and $\sigma_{i}$ is the absorption cross section of a state with momentum $p$ by single PBH.
Summing over all possible particle species and integrating over the phase space in the emission rate Eq.~(\ref{eq:emission rate}), mass loss rate is obtained as follows~\cite{Cheek2022}
\begin{equation}
\begin{aligned}[b]
   \frac{dM_\text{BH}}{dt} &=  \sum_i \left(\frac{dM_\text{BH}}{dt} \right)_i = - \sum_i \int_0^\infty E_i \frac{d^2N_i}{dp \ dt} dp \\
    &= - \frac{27 M_\text{BH}^2}{128 \pi^3 M_p^4} T_\text{BH}^4 \sum_i g_i \int_{z_i}^\infty \frac{\Psi_{i}(x) (x^2 - z_i^2)}{e^x \pm 1} x  dx = - \varepsilon(M_\text{BH}) \frac{ \left( 8\pi M^2_p \right)^2 }{M_\text{BH}^2},
\end{aligned}
\end{equation}
where $x=E/T_\text{BH}$, $z_i = \mu_i/T_\text{BH}$ and $\varepsilon(M_\text{BH}) \equiv \sum_i g_i \varepsilon_i (z_i)$  is an evaporation function from those of each species $\varepsilon_i (z_i)$ given by~\cite{Cheek2022} 
\begin{equation}\label{eq:evap Func}
    \varepsilon_i (z_i) = - \frac{27}{128 \pi^3} \int_{z_i}^\infty \frac{\Psi_{i}(x) (x^2 - z_i^2)}{e^x \pm 1} x dx,
\end{equation}
where $\Psi_{i}(E) = \sigma_{i}(E)/(27 \pi G^2 M_{BH}^2)$. 
Most energy of PBH is transferred during the final stages of the evaporation process, which occurs when the mass of black hole becomes small.
When the produced particle mass is much smaller than the Hawking temperature $\mu_i \ll T_{\rm BH}$ and the geometrical optics limit is valid as $\sigma_{i}(E)|_{\rm GO}= 27 \pi G^2 M_{\rm BH}^2$~\cite{Page1976, Page1977, MacGibbon1990, MacGibbon1991}, where $\Psi_{i}(x) = 1$, the mass loss rate can be approximated as 
\begin{equation}\label{eq:mass loss}
    \frac{d M_\text{BH}}{dt} \simeq - \frac{27\pi}{4} \frac{  \ g_*(T_\text{BH})}{480} \frac{M_p^4}{M_\text{BH}^2},
\end{equation}
which agrees with~\cite{Baldes_2020, Cheek2022}~\footnote{The factor of 27/4 is different from some references~\cite{Baumann:2007yr, Fujita2014,Gondolo2020, Masina:2020xhk, Bernal:2020bjf, Morrison:2018xla}, because they neglect the effect of grey-body factor for analytical estimation.}. 
By solving Eq.~(\ref{eq:mass loss}), the PBH mass evolution can be obtained as
\begin{equation}
    M_\text{BH}(t) = M_\text{in}(t_\text{in}) \left(1- \frac{t-t_\text{in}}{\tau}\right)^{1/3},
\end{equation}
with the PBH lifetime given by 
\begin{equation}\label{eq:lifetime}
    \tau = \frac{4}{27} \frac{160 \ M_\text{in}^3}{\pi \ g_*(T_\text{BH}) M_p^4}.
\end{equation}
The evaporation temperature can be defined as the temperature of the plasma right after the evaporation of PBH, using $\rho_\text{rad}(T_\text{ev})= 3M_p^2 H^2(T_\text{ev}) $.
When PBH is subdominant compared to the background plasma density during its evaporation, we can use that $H^2(T_\text{ev}) = \frac{1}{4 \tau^2}$ and find 
\begin{equation}\label{eq:evapTemperature}
\begin{aligned}[b]
    T_\text{ev} |_\text{RD} &\equiv \left( \frac{45 M_p^2}{2 \pi^2 g_*(T_\text{ev})\tau^2}\right)^{1/4}  \simeq 0.447 \left(\frac{g_*^2(T_\text{BH})}{g_*(T_\text{ev})}\right)^{1/4} \left(\frac{M_p^5}{M_\text{in}^3}\right)^{1/2}  \simeq 30 \ \text{GeV} \left(\frac{10^{6} \ \text{g}}{M_\text{in}}\right)^{3/2}, 
\end{aligned}
\end{equation}
where $g_*(T_\text{BH})\simeq g_*(T_\text{ev})\simeq 100$ were used in the last equation.
In the case when PBHs dominate before evaporation,
we can use $H^2(T_\text{ev}) = \frac{4}{9 \tau^2}$ then the evaporation temperature is slightly higher than in the radiation-dominated one~\cite{Masina:2020xhk}
\begin{equation}\label{eq:ev temperaturePBH}
    T_\text{ev} |_\text{EMDE} \simeq \frac{2}{\sqrt{3}} T_\text{ev} |_\text{RD}.
\end{equation}
After the complete evaporation of PBHs, the Universe again is dominated by radiation.

Since the electromagnetic and hadronic particles produced from PBH evaporation can make trouble with the BBN prediction, we therefore require $T_\text{ev} > T_\text{BBN}\simeq 4 \ \text{MeV}$ not to spoil the successful nucleosynthesis~\cite{Baumann:2007yr,Kawasaki:2000en, Hannestad:2004px} . 
This can be translated into an  upper bound on the PBH mass using Eq.~(\ref{eq:evapTemperature}), given by
\begin{equation}\label{eq:PBH mass upper limit}
    \frac{M_\text{in}}{M_p} \lesssim 10.4 \times 10^{13} \left(\frac{g_*^2(T_\text{BH})}{g_*(T_\text{ev})}\right)^{1/6} \quad \Rightarrow \quad M_\text{in}^\text{max} \lesssim 9.7 \times 10^8 \ \text{g}.
\end{equation}
A lower bound on PBH mass comes from CMB observation which puts upper bound on the Hubble rate $H < 2.5 \times 10^{-5} M_p$~\cite{Planck:2018jri}, and the viable black hole should form after that. From Eq.~(\ref{eq:MassPBH}), we get
\begin{equation}\label{eq:PBH mass lower limit}
    M_\text{in} > \frac{4 \pi \gamma  M_p}{2.5 \times 10^{-5}} \simeq \left(\frac{\gamma}{0.2}\right) 0.4 \ \text{g} \quad \Rightarrow \quad M_\text{in}^\text{min} \gtrsim 0.4 \ \text{g}.
\end{equation}

The initial energy density of PBHs normalized to the radiation energy density at the time of formation is characterized by the dimensionless parameter $\beta$ as
\begin{equation}
    \beta\equiv 
    \frac{\rho_{\rm BH}(T_{\rm in})}{\rho_{\rm rad}(T_{\rm in})}.
\end{equation}
 Since $\rho_{\rm BH}\sim a^{-3}$ and $\rho_{\rm rad}\sim a^{-4}$, this ratio grows with the expansion of the Universe until PBH evaporation. Therefore, an initially radiation-dominated Universe will eventually become matter-dominated if the PBHs are still around with large enough $\beta$. The condition 
 of PBH evaporation during radiation domination 
 can be expressed as~\cite{Masina:2020xhk}
\begin{equation}
\begin{aligned}[b]
    \beta \lesssim \frac{\left.T_{\rm ev}\right|_{RD}}{T_{\rm in}} &= \left(\frac{27}{5120}\right)^{1/2} \left(\frac{g_*(T_{\rm BH})}{\gamma}\right)^{1/2} \frac{M_p}{M_{\rm in}} \\
    & \simeq 7 \times 10^{-6} \left(\frac{0.2}{\gamma}\right)^{1/2} \left(\frac{g_*(T_{\rm BH})}{100}\right)^{1/2}\left(\frac{\rm g}{M_{\rm in}}\right).
\end{aligned}
\end{equation}

We note that large $\beta$ is also constrained by the energy density of the gravitational waves (GWs) produced from the PBH evaporation. If we require that the energy density of GWs do not exceed the radiation during BBN, the constraint can be expressed as~\cite{Domenech:2020ssp}
\begin{equation}\label{eq:beta_bound}
    \beta < 1.1 \times 10^{-6} \left(\frac{0.2}{\gamma}\right)^{1/2} \left(\frac{g_*(T_{\rm BH})}{108}\right)^{17/48} \left(\frac{g_*(T_{\rm ev})}{106.75}\right)^{1/16} \left(\frac{10^4 \ \rm g}{M_{\rm in}}\right)^{17/24}.
\end{equation}

In the following sections, we consider these constraints on PBH mass $0.4 \ {\rm g} \lesssim M_{\rm in} \lesssim 9.7 \times 10^8 \ {\rm g}$ from \eq{eq:PBH mass upper limit} and \eq{eq:PBH mass lower limit}, together with the upper limit on the initial PBH energy density ratio $\beta$ given by \eq{eq:beta_bound}. Those limits are illustrated in the Figs. \ref{fig:betaVSMin} and \ref{fig:sB} in the Section \ref{section3}.

\section{Non-thermal WIMPy Baryogenesis}\label{section3}

The authors in Ref.~\cite{CHOI2018657} studied non-thermal WIMPy baryogenesis considering  additional production of WIMP dark matter from decay of heavy particles.
In this paper, we apply this model to the DMs produced from the PBH evaporation. In this section, we summarize the relevant Boltzmann equations and show the main results.

The presence of PBH alongside the standard radiation bath contributes to the additional density and modifies the Hubble expansion rate $H$ of the early Universe as
\begin{equation}
    H^2 = \frac{1}{3 M_p^2} (\rho_r + \rho_\text{BH}).
\end{equation}
Using the comoving energy density and number density of each field, defined by
\dis{
\tilde{\rho}_r \equiv a^4 \times \rho_r, \qquad \tilde{\rho}_{\rm BH} \equiv a^3 \times \rho_{\rm BH}, \qquad \tilde{n}_\chi \equiv a^3 \times n_\chi,
}
with a scale factor $a$, the Boltzmann equations that govern the evolution of the Universe are given by~\cite{Masina:2020xhk, Giudice:2000ex, Bernal:2020bjf, JyotiDas:2021shi, Barman:2021ost}: 
\begin{equation}
\begin{aligned}[b]
    & \frac{dM_\text{BH}}{d \ln(a)} = - \frac{\varepsilon(M_\text{BH})}{H} \frac{(8 \pi M_p^2)^2}{M_\text{BH}^2}, \\
    & \frac{d \tilde{\rho}_\text{BH}}{d \ln(a)} =  \frac{\tilde{\rho}_\text{BH}}{M_\text{BH}} \frac{dM_\text{BH}}{d \ln(a)}, \\
    & \frac{d \tilde{\rho}_r}{d \ln(a)} = -\frac{\varepsilon_\text{SM}(M_\text{BH})}{\varepsilon(M_\text{BH})} \frac{a \ \tilde{\rho}_\text{BH}}{M_\text{BH}} \frac{dM_\text{BH}}{d \ln(a)}, \\
     & \frac{d \tilde{n}_\chi}{d \ln(a)} =  \frac{\tilde{\rho}_\text{BH}}{M_\text{BH}} \frac{\Gamma_{\text{BH}\rightarrow \chi}}{H} - \left\langle\sigma_a \upsilon\right\rangle \frac{(\tilde{n}_\chi^2 - \tilde{n}_{\chi,\text{eq}}^2)}{a^3 H},
     \label{Bolteq}
\end{aligned}
\end{equation}
where $\left\langle\sigma_a \upsilon\right\rangle$ is the thermal averaged total annihilation cross section of WIMP DM and $
\varepsilon_\text{SM}\equiv g_{\rm SM}\sum_i \varepsilon_i (z_i)$ contains only the SM contribution into the evaporation excluding DM.
Additionally, $\Gamma_{\text{BH}\rightarrow \chi}$ is the momentum integrated emission rate of non-thermal DM from PBH evaporation given  by
\begin{equation}
    \begin{aligned}[b]
        \Gamma_{\text{BH}\rightarrow \chi} = \int dp \frac{d^2N_\chi}{dp \ dt} =\frac{27 g_\chi}{128 \pi^3} \frac{M_p^2}{M_\text{BH}}\int_{z}^\infty \frac{\Psi_{\chi}(x) (x^2 - z^2)}{e^x \pm 1}dx,
    \end{aligned}
\end{equation}
where $x=E/T_\text{BH}$, and $z = \ m_\text{DM}/T_\text{BH}$. 
In the geometrical optics limit, the integration of the total emission rate can be solved by analytically and gives~\cite{Cheek2022, Perez-Gonzalez:2020vnz}
\begin{equation}
    \Gamma_{\text{BH}\rightarrow \chi} = \frac{27 g_\chi}{64 \pi^3} \frac{M_p^2}{M_\text{BH}}[-z \ Li_2(-e^{-z}) - Li_3(-e^{-z})],
\end{equation}
where $Li_n$ is the polylog functions of order n. When temperature is higher than DM mass $z\ll 1$, the bracket of the polylog function approaches 0.9. However,  in our numerical calculation, we include the complete greybody factors using a publicly available python package~\cite{Granelli:2020pim}.

The DM number density $n_{\chi,\text{eq}}$ at the thermal equilibrium is given by
\begin{equation}
    n_{\chi,\text{eq}} = \frac{g_\chi}{2 \pi^2} m_\text{DM}^2 T K_2(m_\text{DM}/T),
\end{equation}
where $g_\chi$ is the dark matter degrees of freedom, which is $2$ for spin 1/2 fermionic particle.  In principle, the Boltzmann equation for the energy density of radiation should also contain the contribution from the annihilation of the thermal WIMP DM. However, this contribution is negligible compared to the relativistic density and can be safely ignored. 

For simplicity, we assume a monochromatic mass function for PBHs, implying that all PBHs have identical mass. However, in general, it is more realistic to consider a population of PBHs with a range of different masses. In this case, we can modify the Boltzmann equations to include the mass function and solve them. The result should depend on the mass spectrum and might have more freedom to explain the dark matter abundance and BAU. Additionally, the PBHs are assumed to be of Schwarzschild type without any spin and charge.
We leave those extensions of PBH mass spectrum and properties for future work.

 When PBHs are the dominant component of the Universe, the approximate solutions for the energy density of the PBHs and radiation can be expressed as 
\begin{equation}\label{eq:PBH energy density}
    \rho_{\rm BH}(T) = \rho_{\rm BH}(T_{\rm in}) \left(\frac{a_i}{a}\right)^3 \exp[{-\varepsilon(M_{\rm BH})64 \pi^2 M_p^4/M_{\rm BH}^3 (T-T_{\rm in})}] \propto a^{-3},
\end{equation}
\begin{equation}\label{eq:SM energy density}
    \rho_R(T) \simeq \frac{2 \sqrt{3}}{5} \varepsilon_{\rm SM} \sqrt{\rho_{\rm BH}(T_{\rm in})} \frac{64 \pi^2 M_p^5}{M_{\rm BH}^3} \left(\frac{a_i}{a}\right)^{3/2} \left[1-\left(\frac{a_i}{a}\right)^{5/2}\right] \propto a^{-3/2}.
\end{equation}

The DMs initially follow the thermal equilibrium and freeze out at around $T\sim m_{\rm DM}/25$ without PBH. 
However, due to the continuous production from PBH, DMs reach a quasi-stable state where the production from PBH evaporation and the annihilation processes equal to each other. 
During this epoch, the approximate solution for the number density of DM can be found from Eq.~(\ref{Bolteq}) as
\begin{equation}\label{eq:n_DM}
    n_\chi = n_{\bar{\chi}} \simeq \left(\frac{\Gamma_{\rm BH \rightarrow \chi}}{M_{\rm BH} \left\langle\sigma_a \upsilon\right\rangle} \rho_{\rm BH}\right)^{1/2} \propto a^{-3/2}.
\end{equation}

After PBH evaporation completed, there is no more production of non-thermal DMs, and only the DM annihilation is efficient in the radiation-domination. In this epoch, we can solve the Boltzmann equation for the relic abundance of DM, $Y_\chi$,
\begin{equation}
    \frac{d Y_\chi}{Y^2_\chi} = - \int_{t_\text{ev}}^t dt \left\langle\sigma_a \upsilon\right\rangle s,
\end{equation}
where $t^{-1}=2H(T)$, and find a simple solution for the DM abundance as
\begin{equation}
    Y_\chi(T) = \left[\frac{1}{Y_\chi(T_\text{ev})} - \left\langle\sigma_a \upsilon\right\rangle \left(\frac{s(T)}{2 H(T)} -\frac{s(T_\text{ev})}{2 H(T_\text{ev})} \right)  \right]^{-1}.
\end{equation}
If the abundance $Y_\chi(T_\text{ev})$ from PBH is large enough, then 
the final abundance of DM is only determined by the annihilation cross section given by
\begin{equation}\label{eq:Yann}
    Y_{\chi}^\text{ann}(T_{\rm ev}) \simeq \frac{2 H(T_\text{ev})}{\left\langle\sigma_a \upsilon\right\rangle s(T_\text{ev})} = \sqrt{\frac{45}{2 \pi^2 }}\frac{g_*(T_\text{ev})^{1/2}}{g_{*,s}(T_\text{ev})} \frac{1}{\left\langle\sigma_a \upsilon\right\rangle M_p T_\text{ev}},
\end{equation}
and the corresponding relic density of DM at present is 
\begin{equation}\label{eq:correct relicDM}
\begin{aligned}[b]
    \Omega_\chi^\text{ann} h^2 &\simeq 4.14 \times 10^8 \ \frac{g_*(T_\text{ev})^{1/2}}{g_{*,s}(T_\text{ev})} \frac{1}{\left\langle\sigma_a \upsilon\right\rangle M_p T_\text{ev}}\left(\frac{m_{\rm DM}}{\text{GeV}}\right) \\
      & \simeq 0.115 \ \sqrt{\frac{100}{g_*(T_\text{ev})}}\left(\frac{10^{-8} \ \text{GeV}^{-2}}{\left\langle\sigma_a \upsilon\right\rangle}\right)\left(  \frac{ M_\text{in}}{10^6\text{g}}\right)^{3/2}\left(\frac{m_{\rm DM}}{2 \ \text{TeV}}\right).
\end{aligned}
\end{equation}

\begin{figure}[tbp]
\centering 
\includegraphics[width=.45\textwidth]{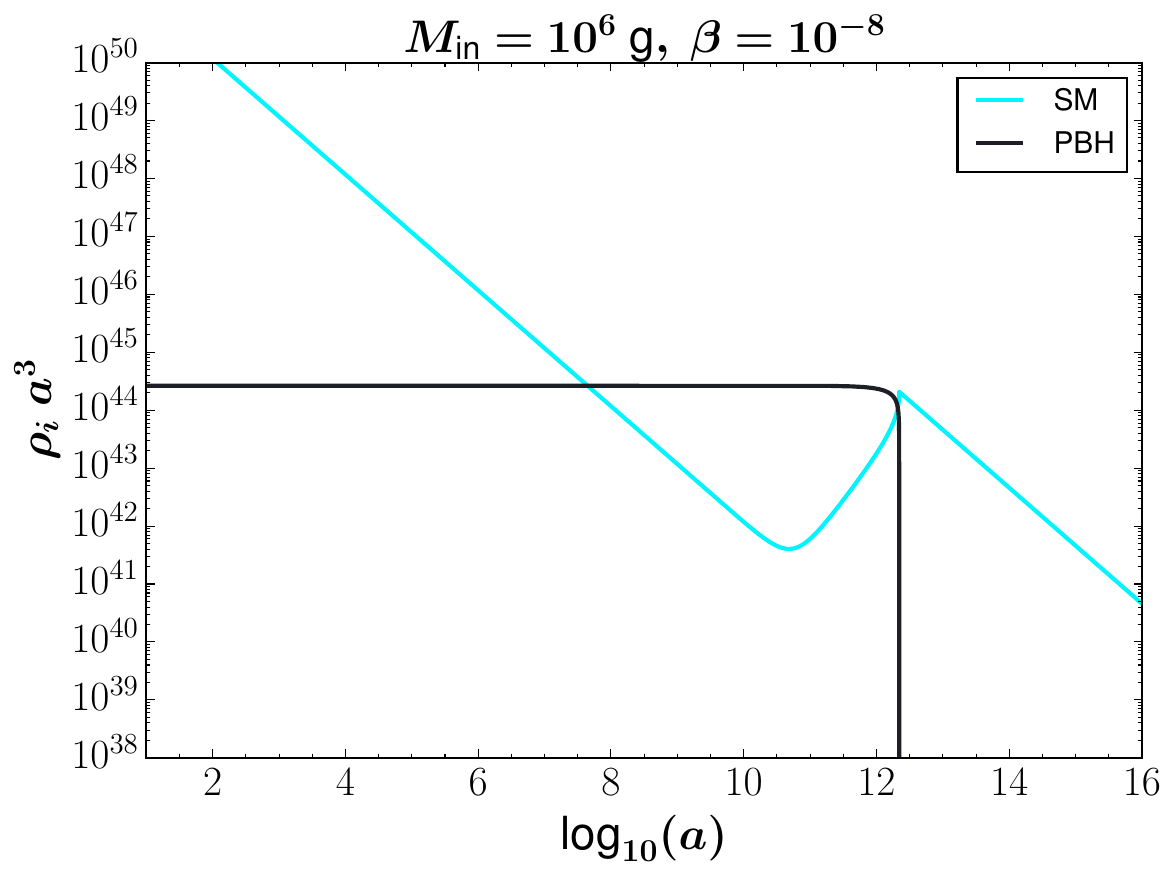}
\hfill
\includegraphics[width=.45\textwidth]{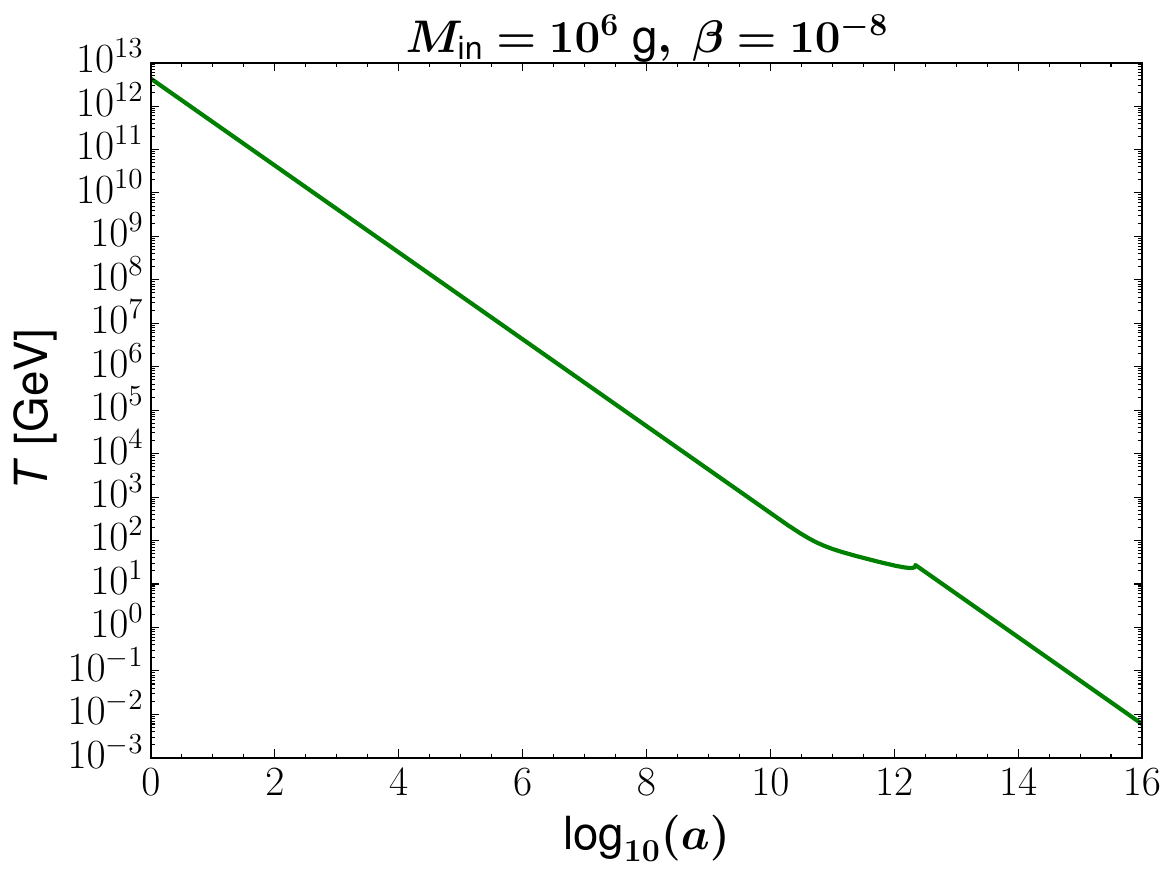}
\hfill
\includegraphics[width=.45\textwidth]{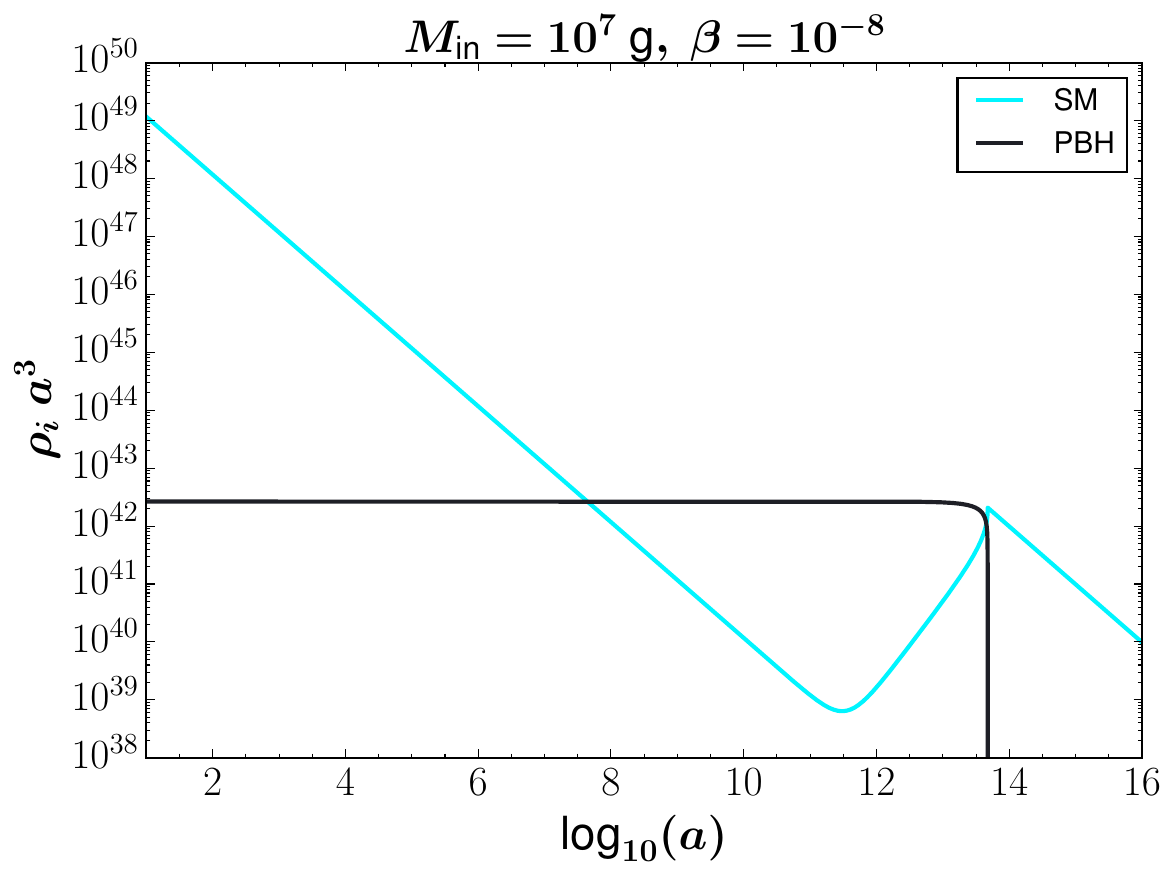}
\hfill
\includegraphics[width=.45\textwidth]{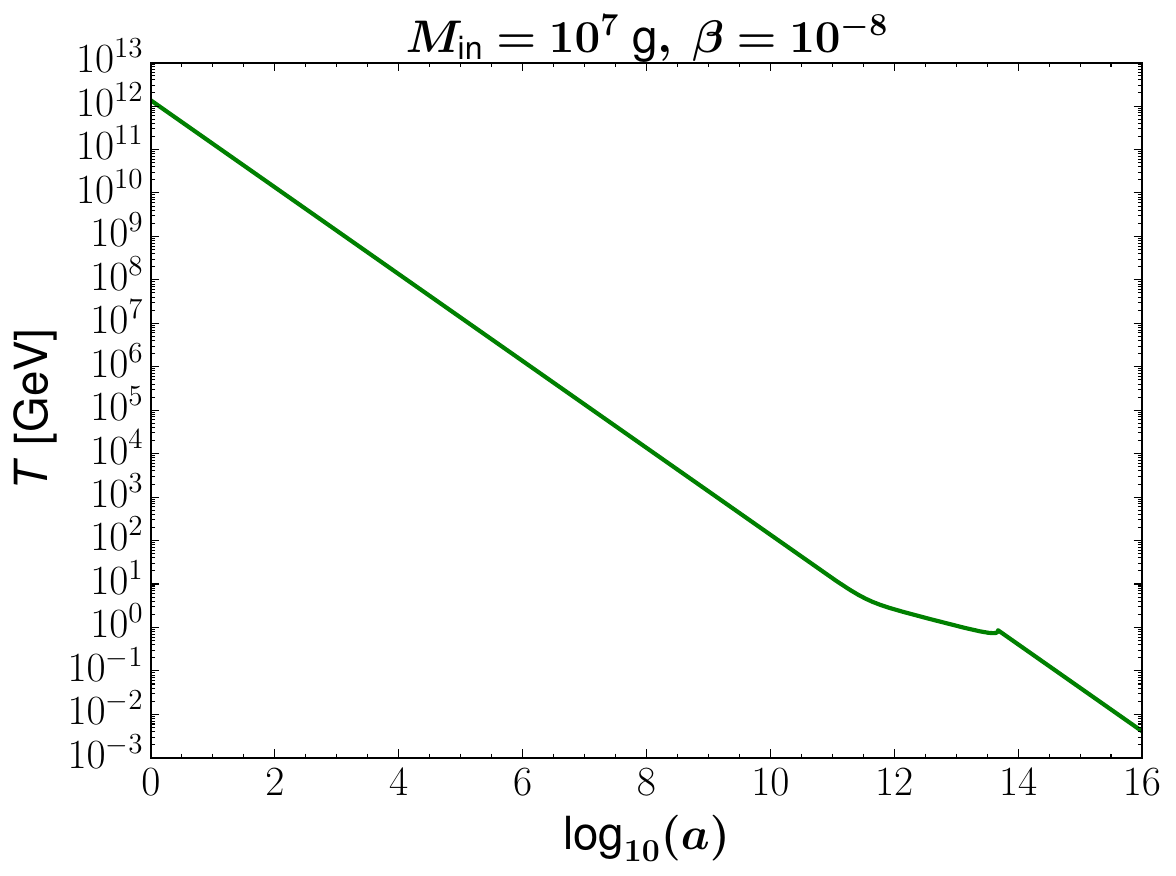}
\caption{\label{fig:EnergyDenandTemperature} Evolution of the energy densities for radiation, PBH (left) and temperature of plasma (right) as a function of scale factor for two different initial mass of PBHs, $M_{\rm in}=10^6$, and $10^7\ \rm{g}$. Here we take $\beta=10^{-8}$.}
\end{figure}

In Fig.~\ref{fig:EnergyDenandTemperature}, we show the numerical results by solving the Boltzmann equations in Eq.~(\ref{Bolteq}). The left panels show the evolution of energy density of radiation (cyan), and PBH (black) as functions of scale factor $a$. With the given initial conditions, PBHs dominate the energy density of the Universe before they evaporate completely, and its energy density shows a sharp decrease at $a\sim 10^{12}$ in the top left panel, for the initial mass of PBHs $M_{\rm in}=10^6\ \rm{g}$ with $\beta=10^{-8}$, and at $a\sim 10^{14}$ in the bottom left panel, for $M_{\rm in}=10^7\ \rm{g}$ with $\beta=10^{-8}$. 
In this EMDE, PBH energy density follows Eq.~(\ref{eq:PBH energy density}), and the radiation density decreases as $a^{-3/2}$ Eq.~(\ref{eq:SM energy density}), so $\tilde{\rho}_{\rm SM}\equiv a^3 \rho_{\rm SM}$ grows by $a^{3/2}$. The right panels show the temperature of thermal bath in terms of the scale factor $a$. A kink in the temperature curve can be observed at the scale factor of approximately $a\sim 10^{12}$ (top) and $10^{14}$ (bottom), which corresponds to the effect of the entropy injection from PBH into thermal plasma. 

\begin{figure}[tbp]
\centering 
\includegraphics[width=.45\textwidth]{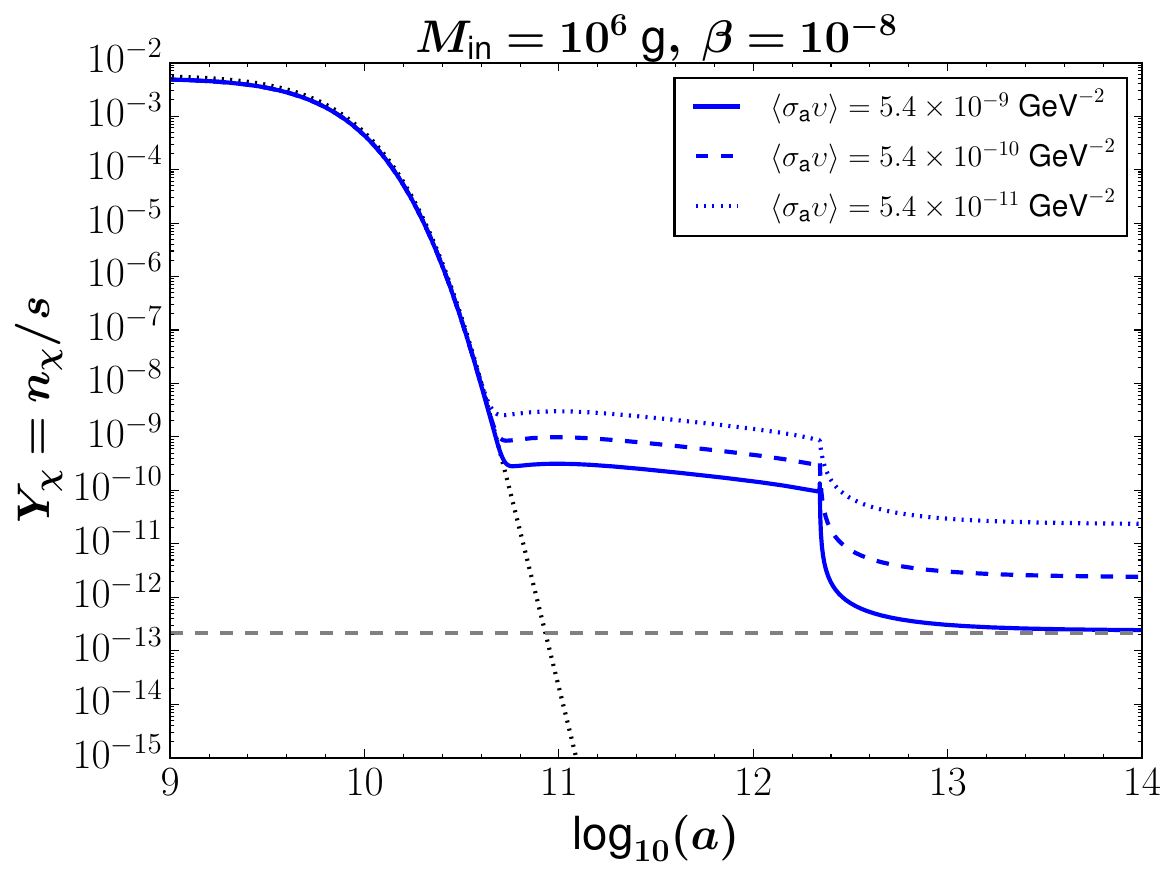}
\hfill
\includegraphics[width=.45\textwidth]{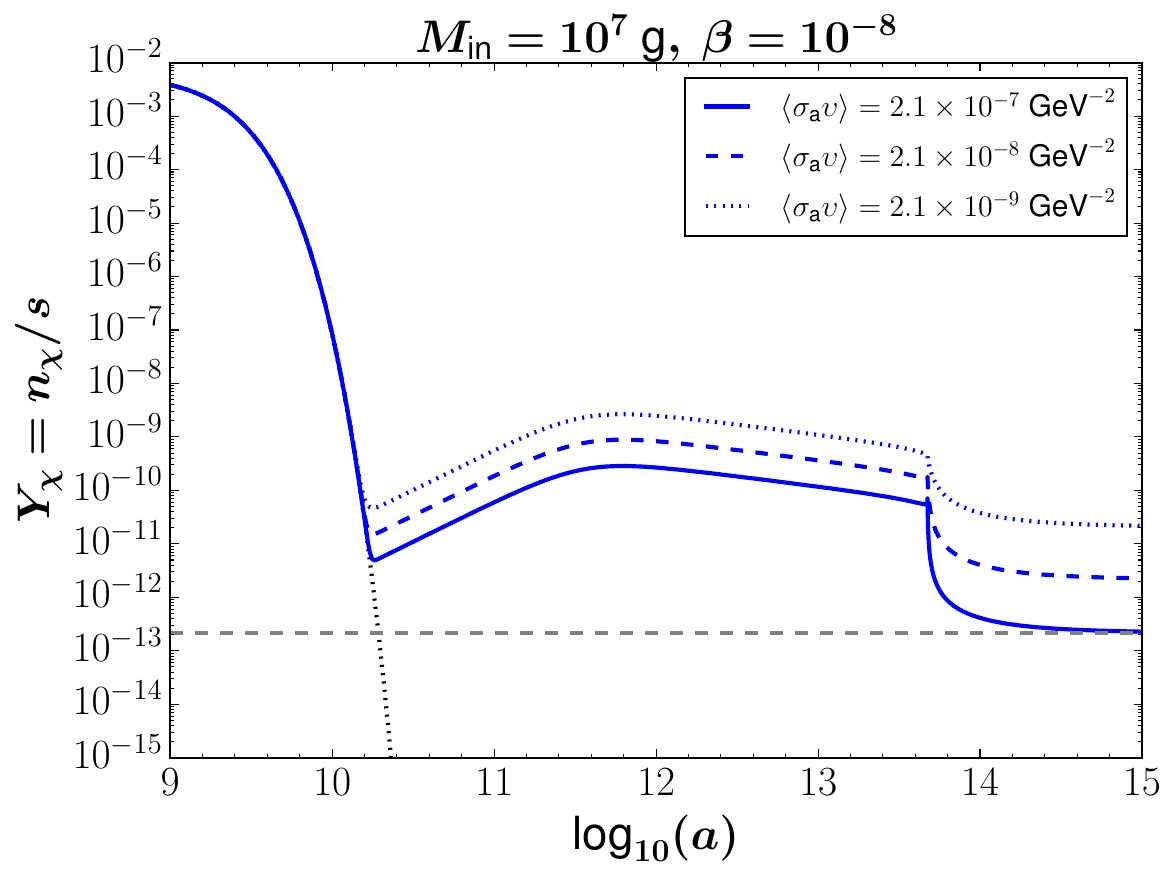}
\caption{\label{fig:YDM} Evolution of DM yield $Y_\chi$
for different total annihilation cross section of DM as mentioned in the plot legend for  two different $M_{\rm in}$ values, $M_{\rm in}=10^6$g (left) and $10^7 \rm g$ (right), respectively. As inputs, we take $\beta=10^{-8}$, $m_{\rm DM} = 2 \ \rm TeV$. 
Here the black dotted line represents the thermal equilibrium of DM, while the grey dashed horizontal line represents the observed correct relic abundance for DM.}
\end{figure}

In Fig.~\ref{fig:YDM}, we show the evolution of the DM yield $Y_\chi$ with blue lines for different total annihilation cross sections of DM.
We plot $Y_\chi$ for two different initial mass $M_{\rm in} = 10^6$g (left) and $10^7$g (right) of PBHs with fixed values of $\beta=10^{-8}$, $m_{\rm DM} = 2 \ \rm TeV$. 
Here the black dotted line represents the thermal equilibrium of DM, while the grey dashed horizontal line represents the observed correct relic abundance for DM of 2 TeV. 
The DMs are initially produced thermally and undergo annihilation around $T_{\rm FO} \simeq 80$ GeV (or scale factor $a\sim 10^{10}$) corresponding to DM $m_{\rm DM} = 2 $ TeV, which is successively followed by the quasi-stable state epoch. During this period, using Eq.(\ref{eq:n_DM}), the DM yield $Y_\chi$ scales as
\begin{equation}\label{eq:YDMscale}
    Y_\chi = \frac{n_\chi + n_{\bar{\chi}}}{s} \sim \frac{a^{-3/2}}{T^3} \propto a^{-3/8},
\end{equation}
where $T^4 \propto a^{-3/2}$ are used in the epoch of EMDE.

After the complete evaporation of PBHs, the non-thermal DMs produced from PBH re-annihilate quickly and become frozen. The final abundance of DM is determined by the total annihilation cross section and the evaporation temperature as shown in Eq.~(\ref{eq:Yann}).
For $M_{\rm in} = 10^6 \ \rm g$ ($M_{\rm in} = 10^7 \ \rm g$), the DM with 2 TeV mass can have correct abundance with $\left\langle\sigma_a \upsilon\right\rangle \simeq 5.4 \times 10^{-9} \ {\rm GeV}^{-2}$ ($\left\langle\sigma_a \upsilon\right\rangle \simeq 2.1 \times 10^{-7} \ {\rm GeV}^{-2}$), respectively.

\begin{figure}[tbp]
\centering 
\includegraphics[width=.45\textwidth]{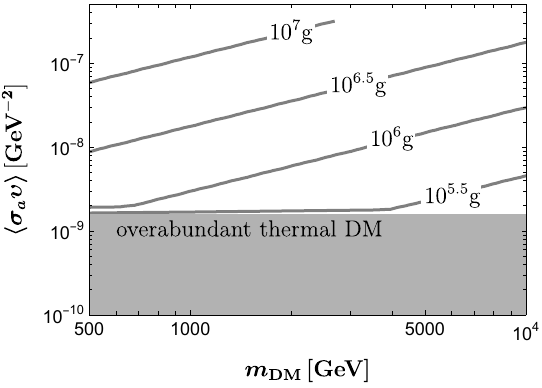}
\hfill
\includegraphics[width=.45\textwidth]{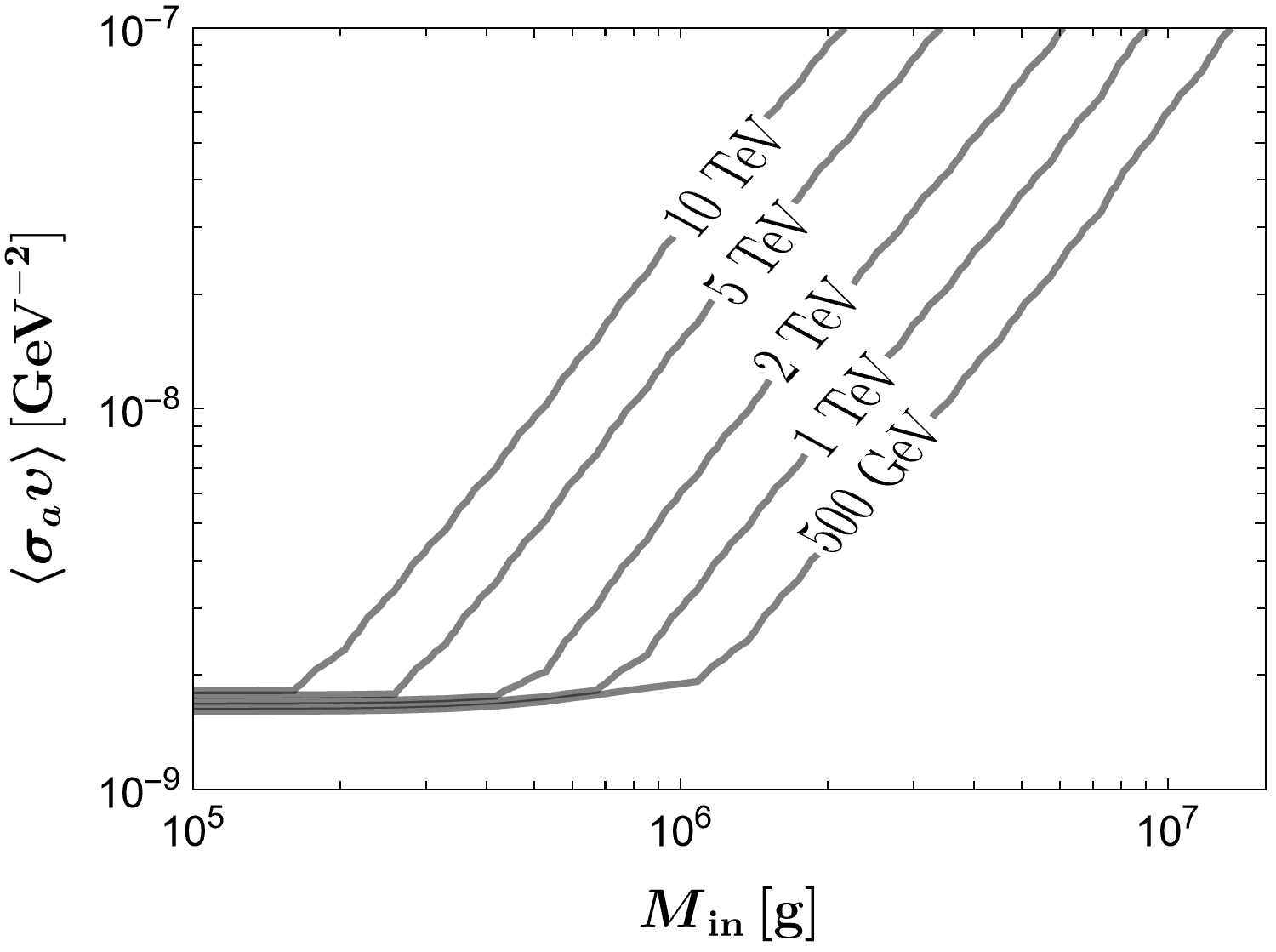}
\caption{ \label{fig:svVSmass} Contour plot for the correct DM relic abundance in the plane of the total DM annihilation cross section and DM mass for several PBH masses (left) and initial mass of PBH for several DM masses (right). When either PBH or DM masses is light, DM  is produced dominantly by the usual thermal freeze-out. Here we take $\beta = 10^{-8}$.
}
\end{figure}

In Fig. \ref{fig:svVSmass}, the left panel shows the total annihilation cross section vs DM mass to produce the correct DM relic abundance  for different initial PBH masses: $M_{\rm in}=10^{5.5}, 10^6, 10^{6.5}$, and $10^7$ g, while the right panel shows PBH mass for  different DM masses: $m_{\rm DM} = 0.5, 1, 2, 5$, and $10$ TeV.  
When the evaporation of PBH occurs before the DM thermal freeze-out,  the relic density is determined by the usual thermal freeze-out, which happens for small DM mass and/or small PBH mass, as appears on the horizontal line.
However, when the evaporation occurs in the PBH dominated epoch, 
larger annihilation cross section is needed for larger PBH mass and larger DM mass. These numerical results are in excellent agreement with the analytic estimation  given by Eq.(\ref{eq:correct relicDM}) within 50\% uncertainty. Note that the result does not depend on $\beta$ for PBH dominated case, though we used $\beta=10^{-8}$ for the numerical calculation.

The DMs from PBH are inherently out-of-equilibrium, satisfying one of the Sakharov's conditions for generating baryon asymmetry. To show the generation of baryon asymmetry in this scenario, for simplicity, we consider an example with baryon number (B) and CP violating annihilation of DM,
with CP asymmetry $\epsilon$ given by
\begin{equation}
    \epsilon = \frac{\sigma_{{B}}(\chi \chi \rightarrow \cdots) - \sigma_{{B}} (\bar{\chi} \bar{\chi} \rightarrow \cdots)}{\sigma_{{B}}(\chi \chi \rightarrow \cdots) + \sigma_{{B}} (\bar{\chi} \bar{\chi} \rightarrow \cdots)}.
\end{equation}
The Boltzmann equation governing the evolution of the baryon asymmetry is written as~\cite{CHOI2018657} 
\begin{equation}\label{eq:BEQ_BA}
\begin{aligned}[b]
    \frac{dn_B}{d t} + 3H n_B = \epsilon \left\langle\sigma_{{B}}\upsilon\right\rangle (n_\chi^2 - n_{\chi,\text{eq}}^2) - \langle\sigma_{\rm wo} v \rangle n_B n_{\psi,\rm eq},
\end{aligned}
\end{equation}
where the second term in the right-hand side is the possible wash-out effect,
and the number density of the exotic quark $\psi$ at equilibrium is $n_{\psi,\text{eq}} = g_\psi/(2 \pi^2) m_\psi^2 T K_2(m_\psi/T)$ with $g_\psi=2$. 
Here we do not consider the specific annihilation modes, and the total cross section is chosen as a free parameter with  $\left\langle\sigma_a\upsilon\right\rangle \geq \left\langle\sigma_{{B}}\upsilon\right\rangle \sim \langle\sigma_{\rm wo} v \rangle $. 
As an example, such concrete models have been studied in the successful WIMPy baryogenesis ~\cite{Cui:2011ab, Bernal:2012gv, Bernal:2013bga, CHOI2018657}, which includes  a vector-like gauge singlet WIMP dark matter $\chi$ and $\bar{\chi}$, singlet pseudoscalars $S_\alpha$, vector-like exotic heavy quark color triplets $\psi_i$ and $\bar{\psi}_i$, and the right-handed up-type SM quark $u$ in the Lagrangian, where the baryon asymmetry is generated via the annihilation $\chi\chi\rightarrow \bar{u}\psi$.

\begin{figure}[tbp]
\centering 
\includegraphics[width=.45\textwidth]{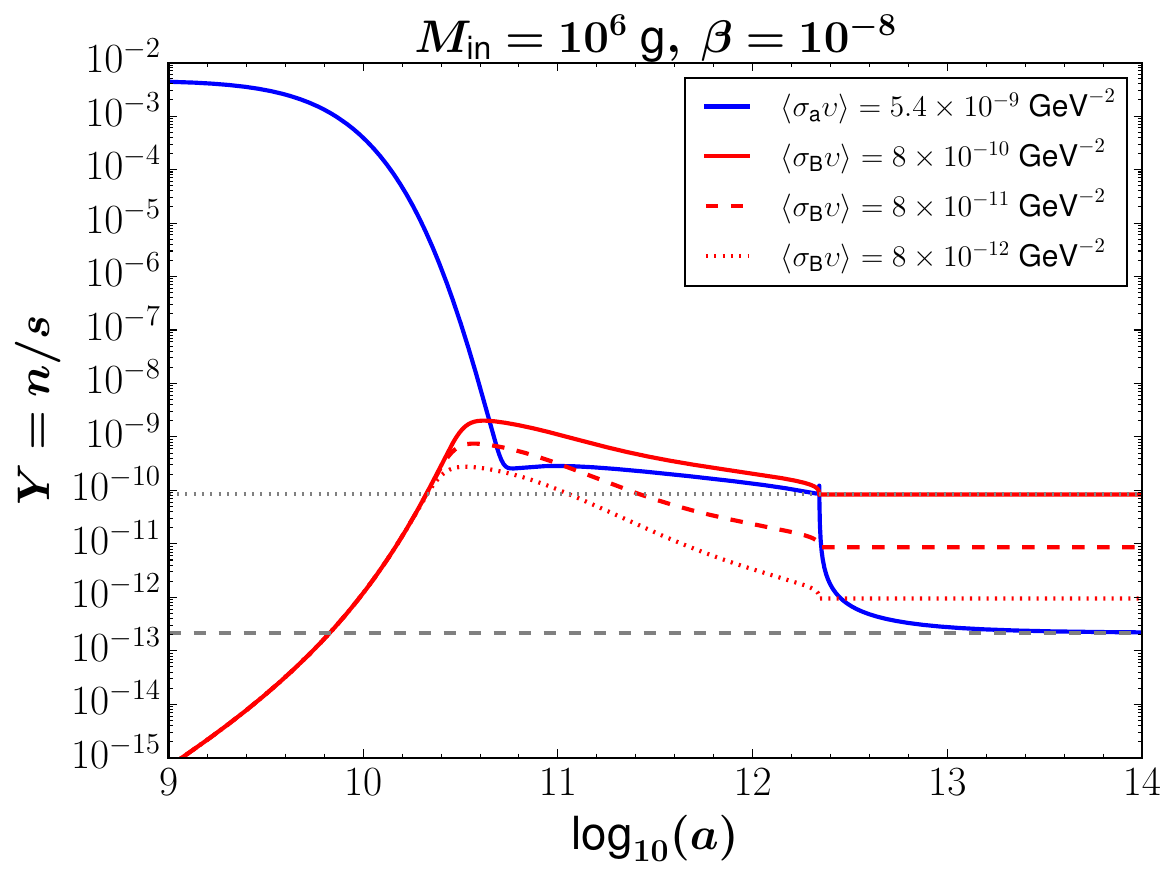}
\hfill
\includegraphics[width=.45\textwidth]{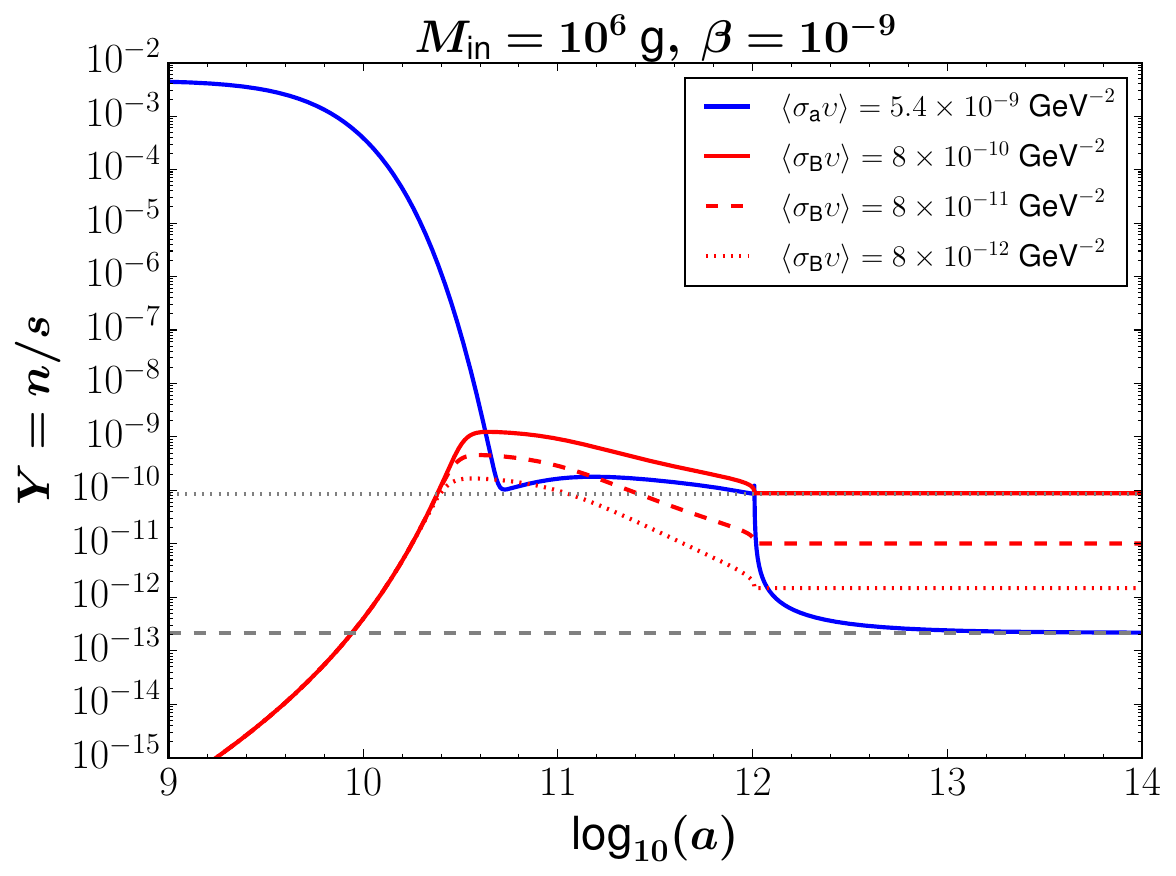} 
\hfill
\includegraphics[width=.45\textwidth]{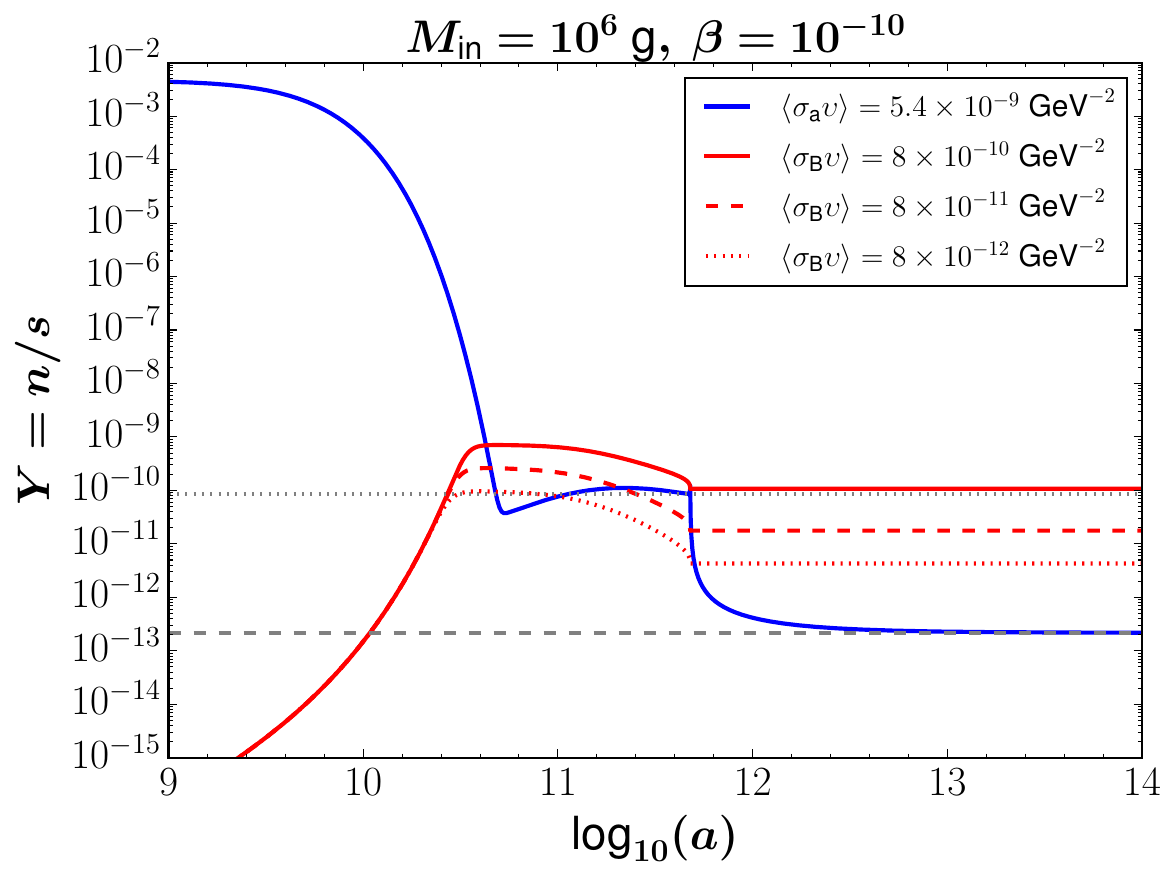} 
\hfill
\includegraphics[width=.45\textwidth]{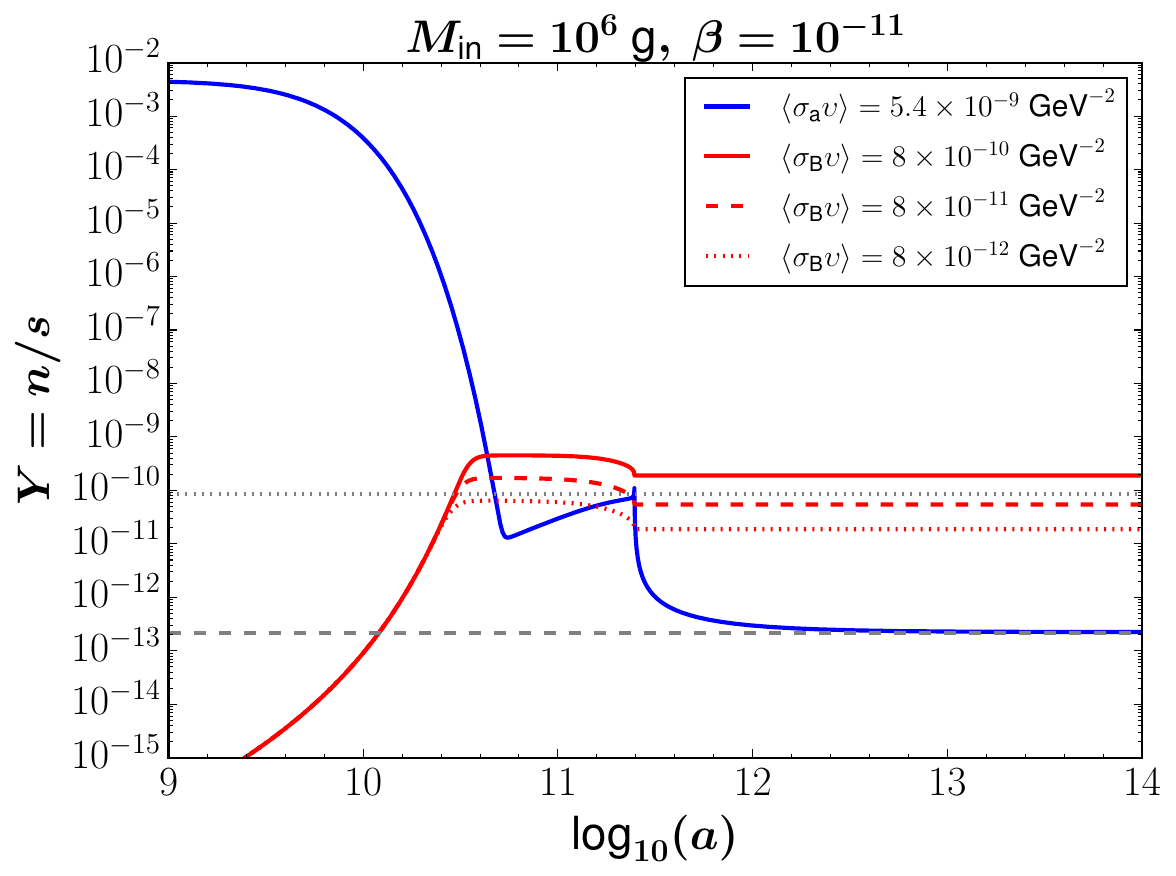} 
\caption{\label{fig:YBA} The evolution of the abundance of baryon asymmetry $Y_B$ 
(red lines) with $M_{\rm in} = 10^6 \rm g$ and different B-violating annihilation cross section $\langle \sigma_B v \rangle$ which is displayed on the legend of each panel, as well as DM yield (blue line). The grey dotted horizontal line shows the observed value of baryon asymmetry, while the grey dashed horizontal line represents the observed correct relic abundance for DM with mass $m_{\rm DM} = 2 \ \rm TeV$. We used $\epsilon = 0.1$, $m_\psi = 3 \ \rm TeV$, and $\langle \sigma_B v \rangle =\langle \sigma_{\rm wo} v \rangle$.}
\end{figure}

During the EMDE by PBH, ignoring the wash-out effect, the approximate scaling solution for baryon asymmetry can be found as ~\cite{CHOI2018657} 
\begin{equation}
    n_B = \frac{2}{3 H} \epsilon \left\langle\sigma_{{B}}\upsilon\right\rangle n_\chi^2 = \frac{2}{\sqrt{3}} \epsilon \Gamma_{\rm BH \rightarrow \chi} \frac{\left\langle\sigma_{{B}}\upsilon\right\rangle}{\left\langle\sigma_a \upsilon\right\rangle} \frac{M_p}{M_{\rm BH}} \sqrt{\rho_{\rm BH}} \propto a^{-3/2},
    \label{nB_scale}
\end{equation}
where we used the result of Eq.~(\ref{eq:n_DM}). 
Now, we can estimate the final abundance of the baryon asymmetry in terms of the PBH number density and the DM annihilation cross section using the DM abundance at the time of $T_{\rm ev}$ as
\begin{equation}\label{eq:Y_B}
\begin{aligned}[b]
     Y_{B} &\equiv \frac{n_B}{s} \simeq \frac{3\sqrt{10}}{2\pi} \epsilon \Gamma_{\rm BH \rightarrow \chi} \frac{\left\langle\sigma_{{B}}\upsilon\right\rangle}{\left\langle\sigma_a \upsilon\right\rangle} \frac{M_p}{M_{\rm BH}}\left(\frac{1}{g_*(T_{\rm ev})}\right)^{1/2} \frac{1}{T_{\rm ev}}\\
     &\simeq 8.7 \times 10^{-11} \bfrac{\epsilon}{0.1}\bfrac{f_\sigma}{0.15}\left(\frac{100}{g_*(T_{\rm ev})}\right)^{1/2}\bfrac{10^6 {\rm g}}{M_{\rm in}}^{1/2}. 
\end{aligned}
\end{equation}
where $f_\sigma = \left\langle\sigma_{{B}}\upsilon\right\rangle/ \left\langle\sigma_a \upsilon\right\rangle$. The baryon asymmetry is inversely proportional to the square root of the initial PBH mass $M_{\rm in}^{1/2}$.

\begin{figure}[tbp]
\centering 
\includegraphics[width=.45\textwidth]{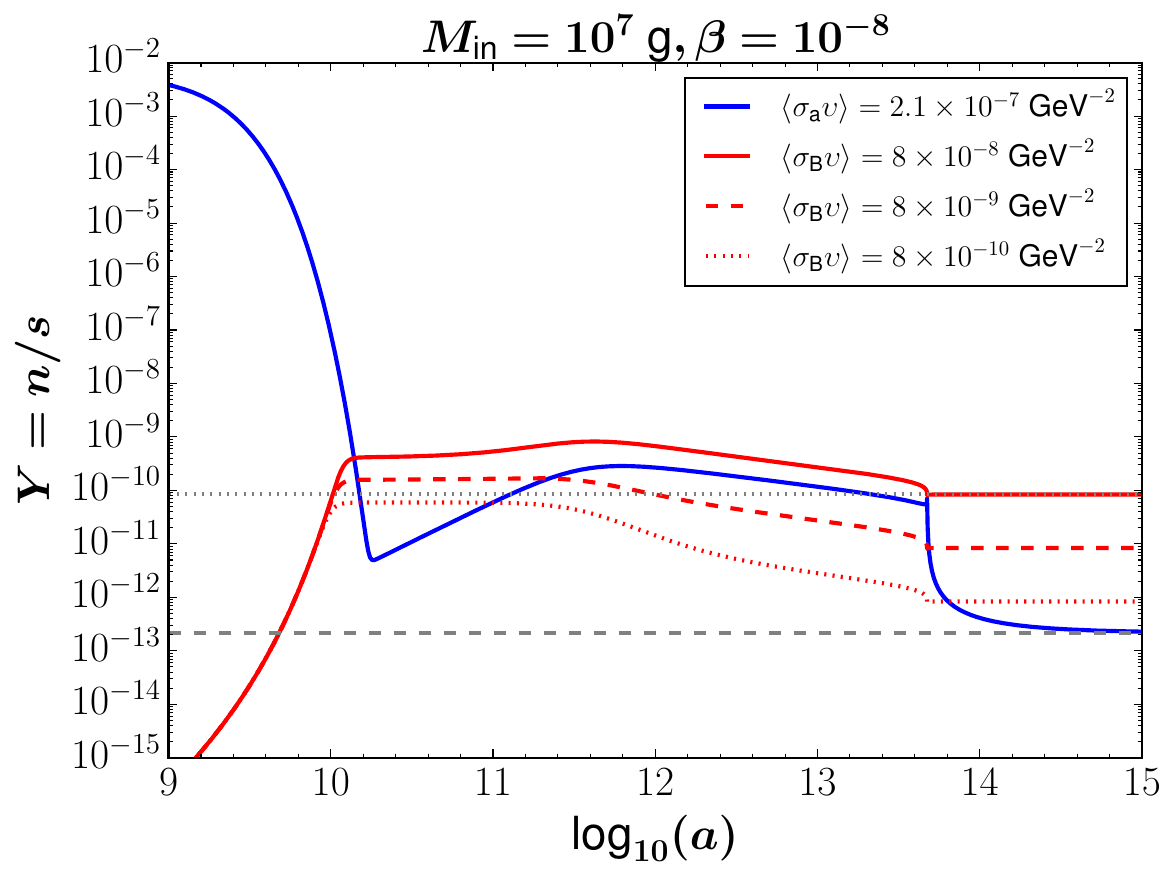}
\hfill
\includegraphics[width=.45\textwidth]{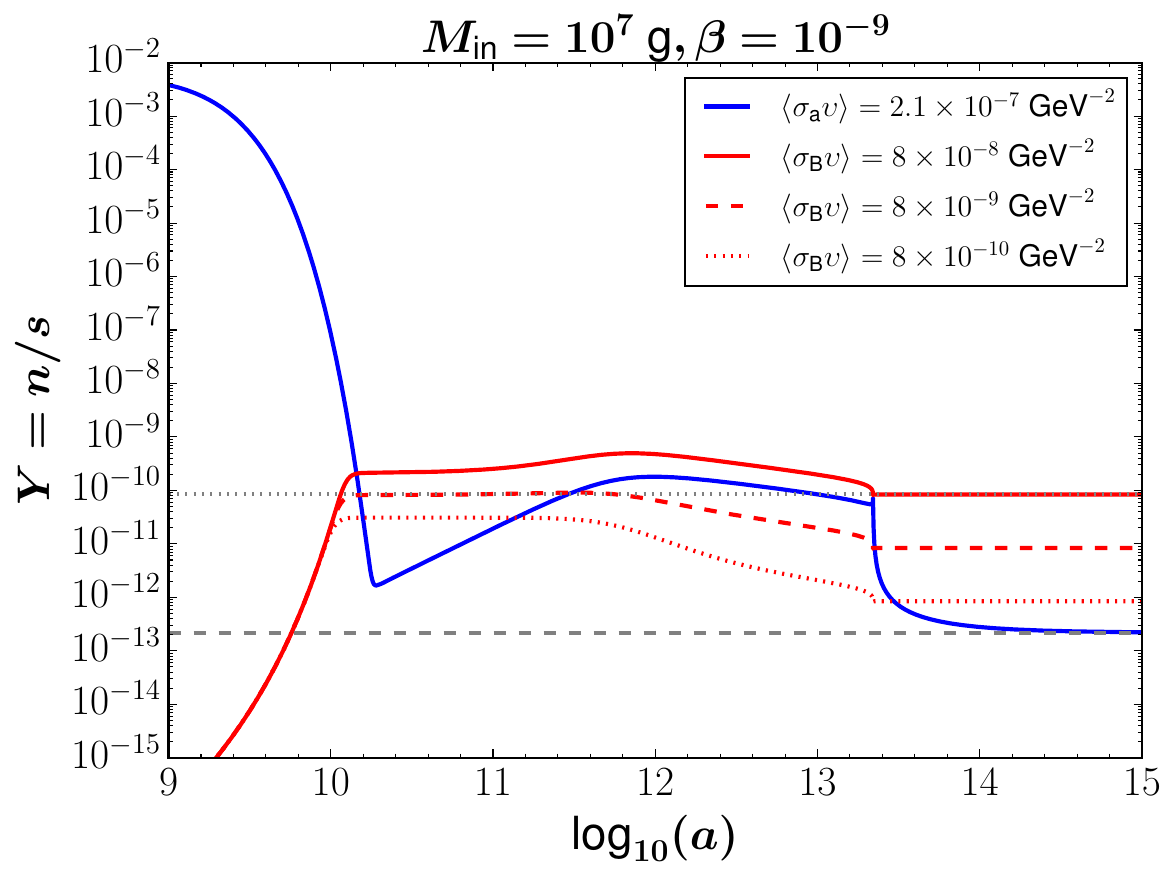} 
\hfill
\includegraphics[width=.45\textwidth]{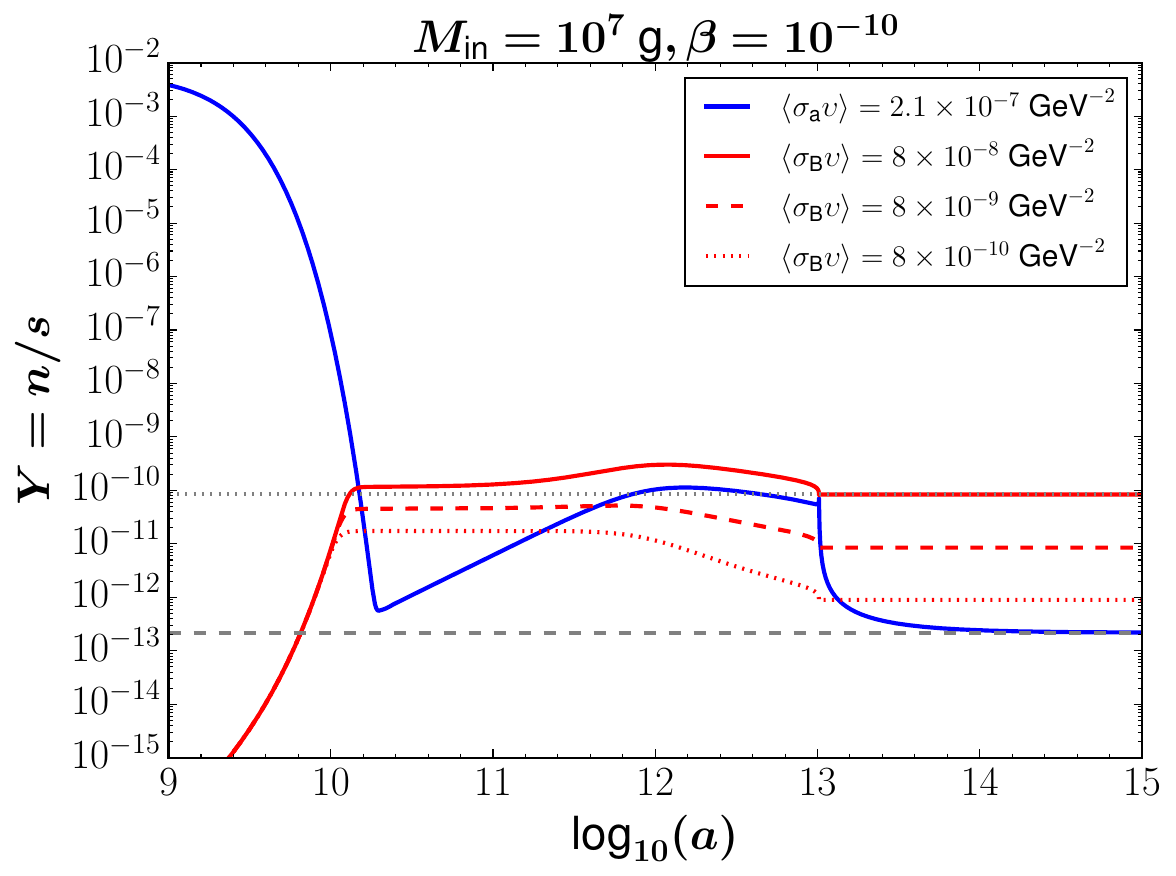} 
\hfill
\includegraphics[width=.45\textwidth]{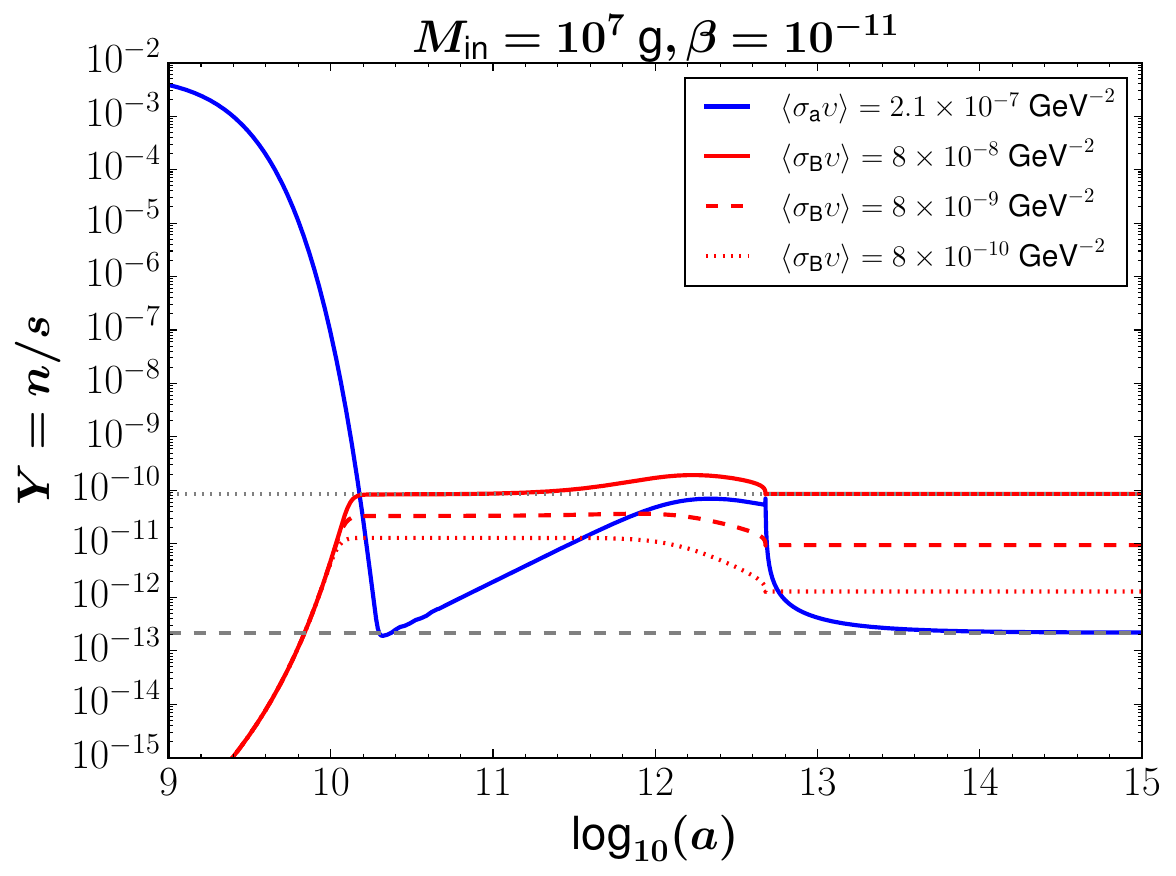} 
\caption{\label{fig:YBA_M7} Same as Fig.~\ref{fig:YBA} for $M_{\rm in} = 10^7 \rm g$. }
\end{figure}

In Fig.~\ref{fig:YBA}, we show the evolution of the baryon asymmetry as well as that of the DM yield with $m_{\rm DM}=2$ TeV and $M_{\rm in} = 10^6 \ \rm g$, for different $\beta$ parameter ($\beta=10^{-8}, 10^{-9}, 10^{-10}$, and $10^{-11}$) and B-violating annihilation cross section $\langle \sigma_B v \rangle = 8\times 10^{-12}, \  8\times 10^{-11}$ and $8\times 10^{-10} \ {\rm GeV}^{-2}$
 with assuming the same wash-out effect $\langle \sigma_{\rm wo} v \rangle = \langle \sigma_B v \rangle$. 
Initially at high temperature, due to the wash-out effect, the generation of baryon asymmetry is suppressed. When the temperature becomes lower than the mass $\psi$, here we use  $m_\psi = 3 \ \rm TeV$, $Y_B$ begins to grow. When the non-thermal DM produced from PBH becomes larger than that from thermal one, $Y_B$ follows the scaling solution estimated in Eq.(\ref{nB_scale}) until the complete evaporation of PBH. 
Finally, after the PBH evaporation, the baryon asymmetry becomes frozen giving the observed relic abundance which is illustrated by the red solid line 
whose cross section is consistent with the semi-analytical estimation Eq.~(\ref{eq:Y_B}). Note that we plot for $Y_B$ taking into account the thermally averaged total annihilation cross section  $\left\langle\sigma_a \upsilon\right\rangle \simeq 5.4 \times 10^{-9} \ {\rm GeV}^{-2}$ (for $M_{\rm in} = 10^6 \ \rm g$) and $\left\langle\sigma_a \upsilon\right\rangle \simeq 2.1 \times 10^{-7} \ {\rm GeV}^{-2}$ (for $M_{\rm in} = 10^7 \ \rm g$) which only give the correct relic abundance for DM. 
From these figures, we can see that the baryon asymmetry is quite independent on the parameter $\beta$. Only for very small $\beta$, where the PBH domination is not guaranteed, the discrepancy begins to appear.
In Fig.~\ref{fig:YBA_M7}, we also show the same figure as in Fig.~\ref{fig:YBA} but for $M_{\rm in} = 10^7 \rm g$.

In our model, the mass of $\psi$ is constrained to be  $T_{\rm ev} < m_\psi < 2 m_{\rm DM}$. The washout effect can be Boltzmann suppressed at  large mass of $\psi$ than PBH evaporation temperature, $T_{\rm ev}<m_{\psi}$. However, $m_\psi$ should be less than $2 m_{\rm DM}$, because B-violating DM annihilation channel is kinematically forbidden for heavier $\psi$. The lower bound for $m_\psi$ can be also found as $2.3 \ \rm TeV$ from the process $pp\rightarrow \tilde{g} \Tilde{g} \rightarrow 4j + \slashed{E}_{T}$ at $95 \%$ CL of the current gluino searches at LHC~\cite{PDG2022}.
Considering these constraints, we used $m_\psi = 3 \ \rm TeV$ in the previous figures as an illustration for DM mass $2$ TeV.


\begin{figure}[tbp]
\centering 
\includegraphics[width=.45\textwidth]{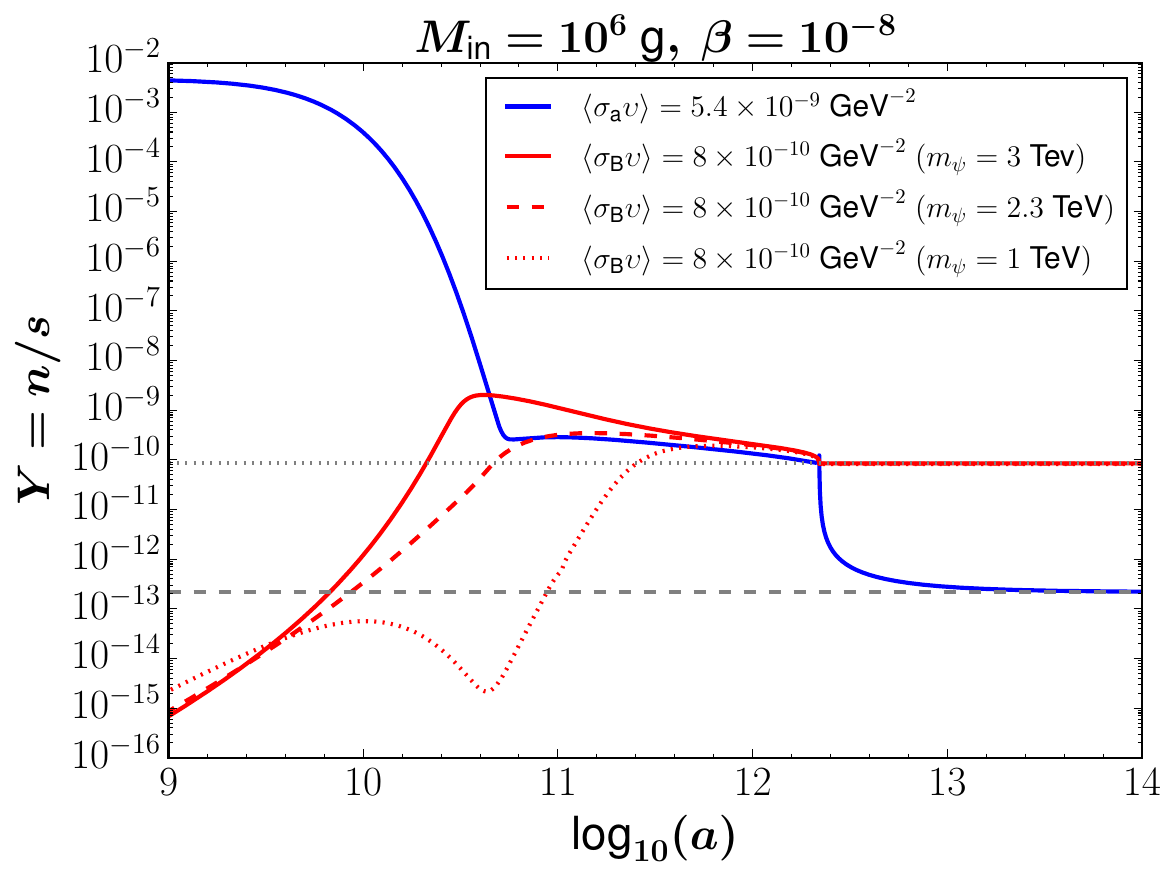}
\hfill
\includegraphics[width=.45\textwidth]{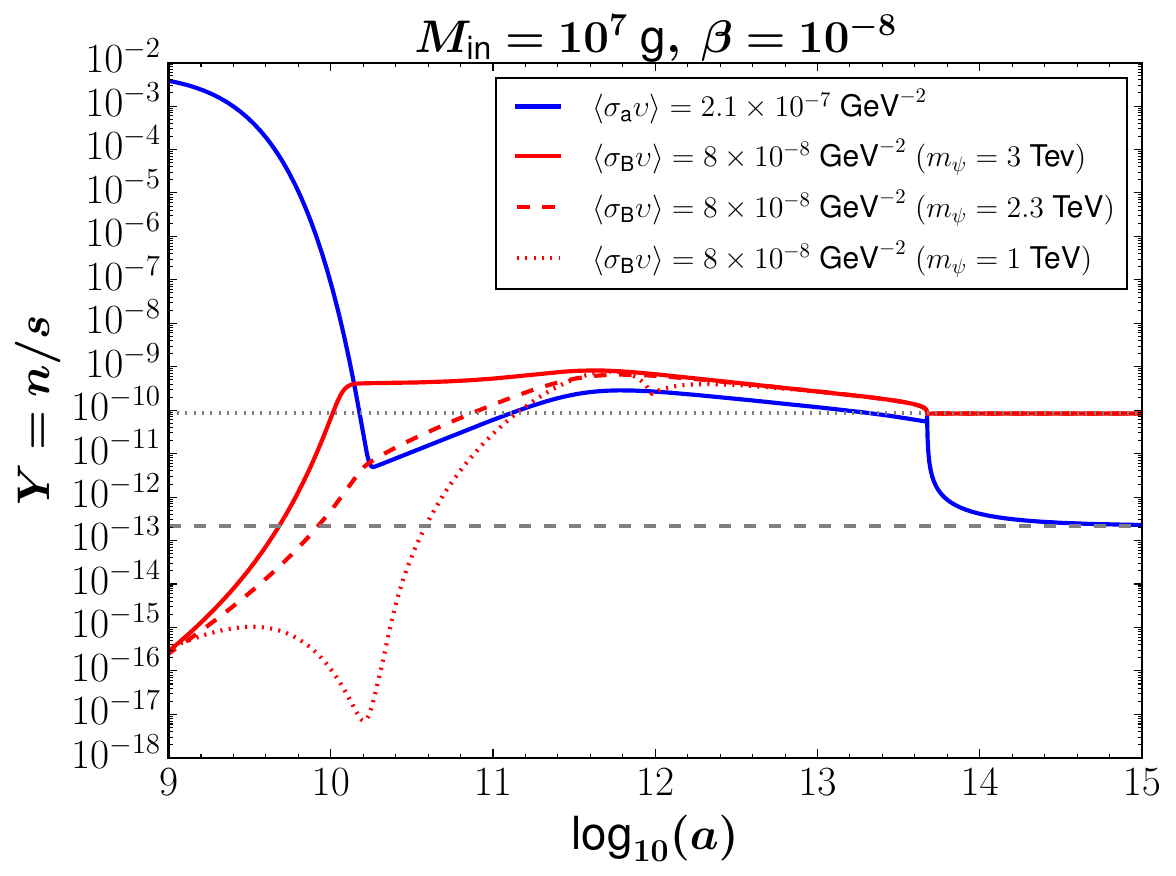}
\caption{\label{fig:Ywith Psimass} Plots of the $Y_{\rm DM}$ and $Y_B$ for different mass values of the exotic quark $\psi$ with $m_\psi = 3 \ \rm TeV$ (solid lines), $m_\psi = 2.3 \ \rm TeV$ (dashed line) and $m_\psi = 1 \ \rm TeV$ (dotted line). Initial parameters are the same as in Fig.~\ref{fig:YBA}. We  used that $\left\langle\sigma_a \upsilon\right\rangle = 5.4 \times 10^{-9} \ {\rm GeV}^{-2}$,   $\left\langle\sigma_B \upsilon\right\rangle \simeq 8 \times 10^{-10} \ {\rm GeV}^{-2}$ for $M_{\rm in} = 10^6 \ \rm g$  (left), and  $\left\langle\sigma_a \upsilon\right\rangle = 2.1 \times 10^{-7} \ {\rm GeV}^{-2}$, $\left\langle\sigma_B \upsilon\right\rangle \simeq 8 \times 10^{-8} \ {\rm GeV}^{-2}$ for  $M_{\rm in} = 10^7 \ \rm g$ (right), respectively.}
\end{figure}

We show the washout effect by changing $m_\psi$ in Fig.~\ref{fig:Ywith Psimass}. Here we used the $m_\psi = 3 \ \rm TeV$ (red solid line), $m_\psi = 2.3 \ \rm TeV$ (red dashed line) and $m_\psi = 1 \ \rm TeV$ (red dotted line) with $m_{\rm DM}=2$ TeV. We can explicitly see that the washout is effective before the PBH evaporation time, afterwards the washout is suppressed and $Y_B$ can be sizeable as giving the correct relic for baryon asymmetry. 
Here, we have adapted the correct cross section values from the results presented in Fig.~\ref{fig:YBA} and \ref{fig:YBA_M7} for the $Y_B$ plots.

\begin{figure}[tbp]  
\centering 
\includegraphics[width=.45\textwidth]{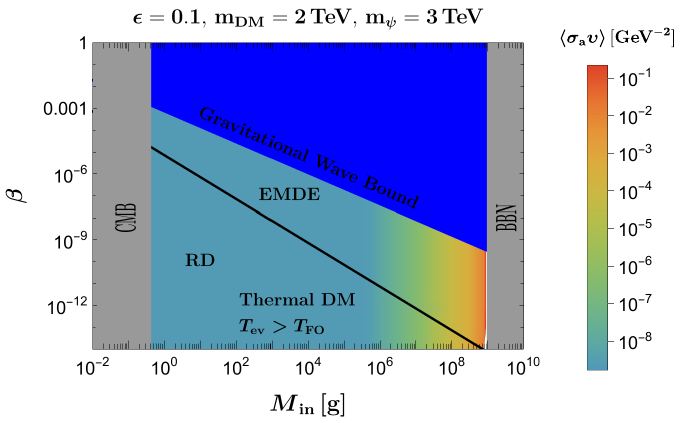}
\hfill
\includegraphics[width=.45\textwidth]{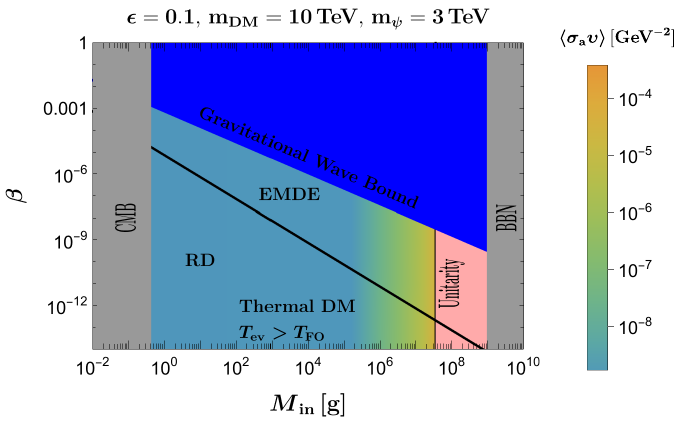}
\caption{ \label{fig:betaVSMin}  Contour plot of $\langle{\sigma_a v}\rangle$ which gives the correct DM relic abundance for $m_{\rm DM}=2$ TeV (left panel) and  $m_{\rm DM}=10$ TeV (right panel) in $(M_{\rm in},\beta )$ parameter space. We used $m_\psi = 3 \ \rm TeV$ and $\epsilon = 0.1$ as before. The constraints on the initial mass for PBH from 
CMB  and BBN are shaded with grey color and the constraint on $\beta$ from GWs with blue color. In the region with dark purple, DMs are produced thermally with $T_{\rm ev}>T_{FO}$. In the right panel, the unitarity bound is shown with pink color. Above the solid black diagonal line, the PBH can dominate the Universe as behaving EMDE before it evaporates completely. 
}
\end{figure}

In Fig.~\ref{fig:betaVSMin}, we show the contour plot of $\langle{\sigma_a v}\rangle$ which could be capable to produce the correct relic abundance for DM with masses $m_{\rm DM}=2$ TeV (left panel) and  $m_{\rm DM}=10$ TeV (right panel) in $(M_{\rm in},\beta )$ parameter space. When $m_{\rm DM}=2$ TeV ($m_{\rm DM}=10$ TeV), thermal DM production becomes dominant  for  $M_{\rm in} \lesssim 10^{5.6} \ \rm{g}$ ($M_{\rm in} \lesssim 10^{5.2} \ \rm{g}$), respectively, which are illustrated by sea green shaded regions. We also show the constraints on the initial mass for PBH with grey shaded regions, from BBN in Eq.~(\ref{eq:PBH mass upper limit}) and from CMB in Eq.~(\ref{eq:PBH mass lower limit}), and the constraint on $\beta$ from GWs in Eq.~(\ref{eq:beta_bound}) with blue color. Above the solid line, the PBHs can dominate the Universe before they evaporate completely.
For $m_{\rm DM} = 10 \ {\rm TeV}$, we also put the unitarity bound on the total annihilation cross section of DM with pink color, which is given for $s$-wave in the non-relativistic regime as~\cite{Kamionkowski},
\begin{equation}
    (\sigma \upsilon_{\rm rel})_{\rm max} = \frac{4 \pi}{m_{\rm DM}^2 \upsilon_{\rm rel}},
\end{equation}
where $\upsilon_{\rm rel}=\sqrt{2 T_{\rm ev}/m_{\rm DM}}$ is the relative velocity between incoming DM particles. Here we used that the non-thermal DM first kinetically thermalized with background particles before reannihilation. From both panels, it is evident that the correct relic abundance for DM can be found to be rather independent of $\beta$ parameter.

\begin{figure}[tbp]  
\centering 
\includegraphics[width=.45\textwidth]{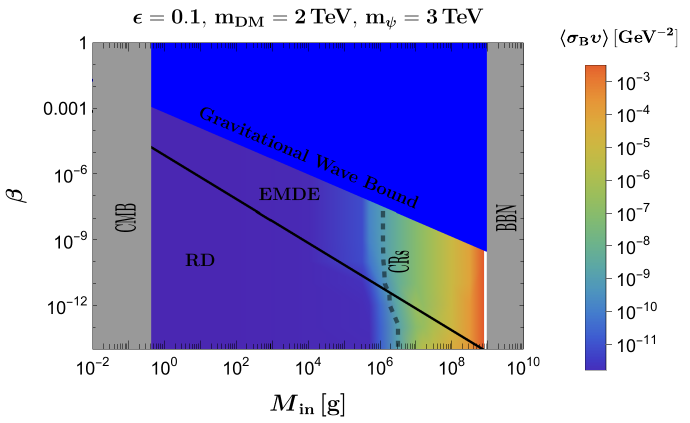}
\hfill
\includegraphics[width=.45\textwidth]{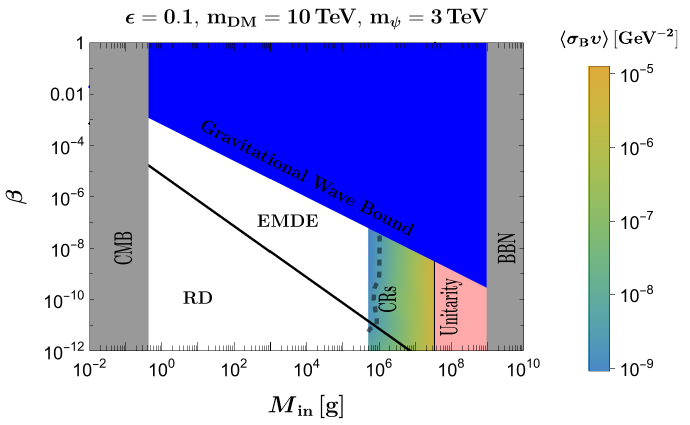}
\hfill
\includegraphics[width=.45\textwidth]{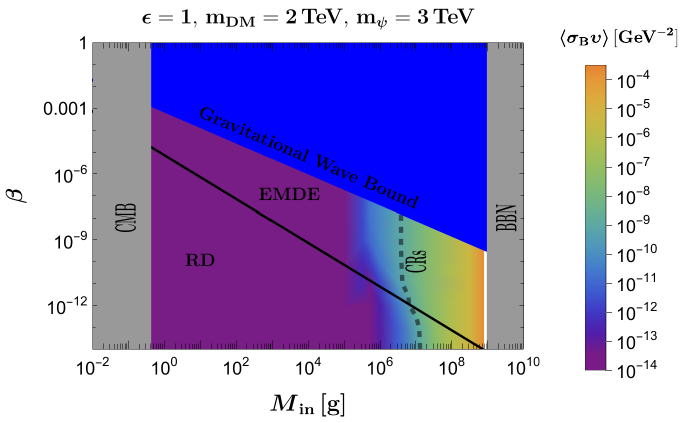}
\hfill
\includegraphics[width=.45\textwidth]{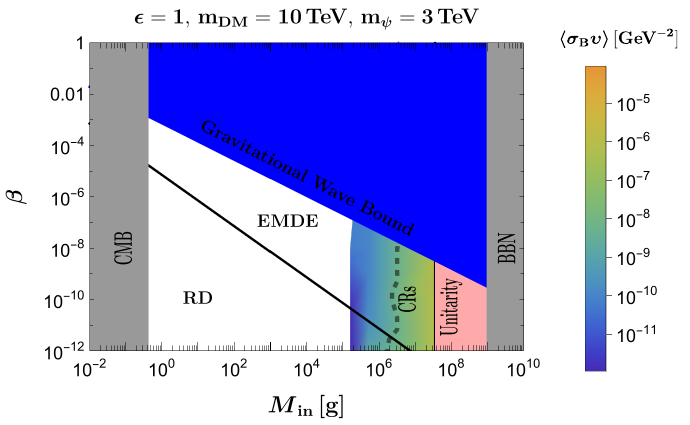}
\caption{ \label{fig:sB}  Contour plot of $\langle{\sigma_B v}\rangle$ which gives the observed baryon asymmetry for $m_{\rm DM}=2$ TeV (left panel) and  $m_{\rm DM}=10$ TeV (right panel) in $(M_{\rm in},\beta )$ parameter space. In the upper panels  $\epsilon=0.1$ is used, while in the lower panels $\epsilon=1$ is used. All constraints, except for cosmic rays from indirect detection, are the same as Fig.~\ref{fig:betaVSMin}. The constraints from indirect detection are shown with dashed line, where the left region is preferred.
}
\end{figure}

In Fig.~\ref{fig:sB}, contour plots of $\langle{\sigma_B v}\rangle$ corresponding to the observed baryon asymmetry $Y_B$ are shown in $(M_{\rm in},\beta )$ parameter space for two different DM masses as previous in Fig.~\ref{fig:betaVSMin}. In the upper panels $\epsilon=0.1$ is used, while in the lower panels $\epsilon=1$ is used. Cosmic ray bounds, are shown with dashed black lines, for B-violating cross section being compatible with the indirect detection experiments of dark matter are found as $\langle \sigma_B v\rangle \simeq 2\times 10^{-9} \ {\rm GeV}$ ($\simeq 5 \times 10^{-9} \ {\rm GeV}$) when $m_{\rm DM}=2$ TeV ($m_{\rm DM}=10$ TeV), respectively.  For $m_{\rm DM}=10 {\rm TeV}$ (right panel), when $M_{\rm in}\lesssim 10^{5.7} {\rm g}$, PBH evaporation completed before DM freeze-out, thus the DMs produced from PBH are thermalized. The baryon asymmetry from thermal WIMP DM annihilation is erased due to wash-out effect, caused by the relatively small $m_\psi=3{\rm TeV}$.

\section{Signals in the indirect detection of dark matter}\label{IDDM}

In our model, the WIMP DM is produced from the evaporation of PBH after thermal freeze-out, and the large total  annihilation cross section is needed to generate correct DM abundance.
As we can see in Fig.~\ref{fig:svVSmass}, the total annihilation cross sections could be much larger than $10^{-9} \  {\rm GeV}^{-2}$ and it could be  $10^{-7} \ {\rm GeV}^{-2}$ for DM with mass 1 TeV and PBH with mass $10^7$ g.
If these large annihilation may produce charged particles, the DMs should have been observed already in the indirect detection experiments. Considering the absence of signals in the indirect detection of DM, the main annihilation products of DM must either be neutrinos or hidden particles, for which the upper limit exceeds approximately $10^{-7} \ {\rm GeV}^{-2} $ with DM mass above TeV~\cite{Arguelles:2019ouk}.

The successful generation of baryon asymmetry in our model needs baryon number violating annihilation of dark matter. The annihilation cross section for that is around $8\times 10^{-10} \ {\rm GeV}^{-2}$ and $8\times 10^{-8} \ {\rm GeV}^{-2}$
for PBH mass $10^6$g and $10^7$g, respectively, for 2 TeV DM, which can be probed in the indirect detection of DM.

There are robust bounds on DM annihilation into baryons from \textit{Planck} measurement of the CMB ~\cite{Planck:2018vyg}, the Alpha Magnetic Spectrometer (AMS-02) measurements of cosmic rays~\cite{Cuoco:2017rxb, Cholis:2019ejx} and \textit{Fermi} Large Area Telescope (Fermi-LAT) of gamma rays~\cite{Fermi-LAT:2016afa, Abazajian:2014fta, Fermi-LAT:2016uux, Fermi-LAT:2015att} 
and IceCube neutrino telescope~\cite{IceCube:2016oqp, IceCube:2017rdn, Arguelles:2019ouk} up to TeV mass scale for DM. 

\begin{figure}[tbp]
\centering 
\includegraphics[width=.45\textwidth]{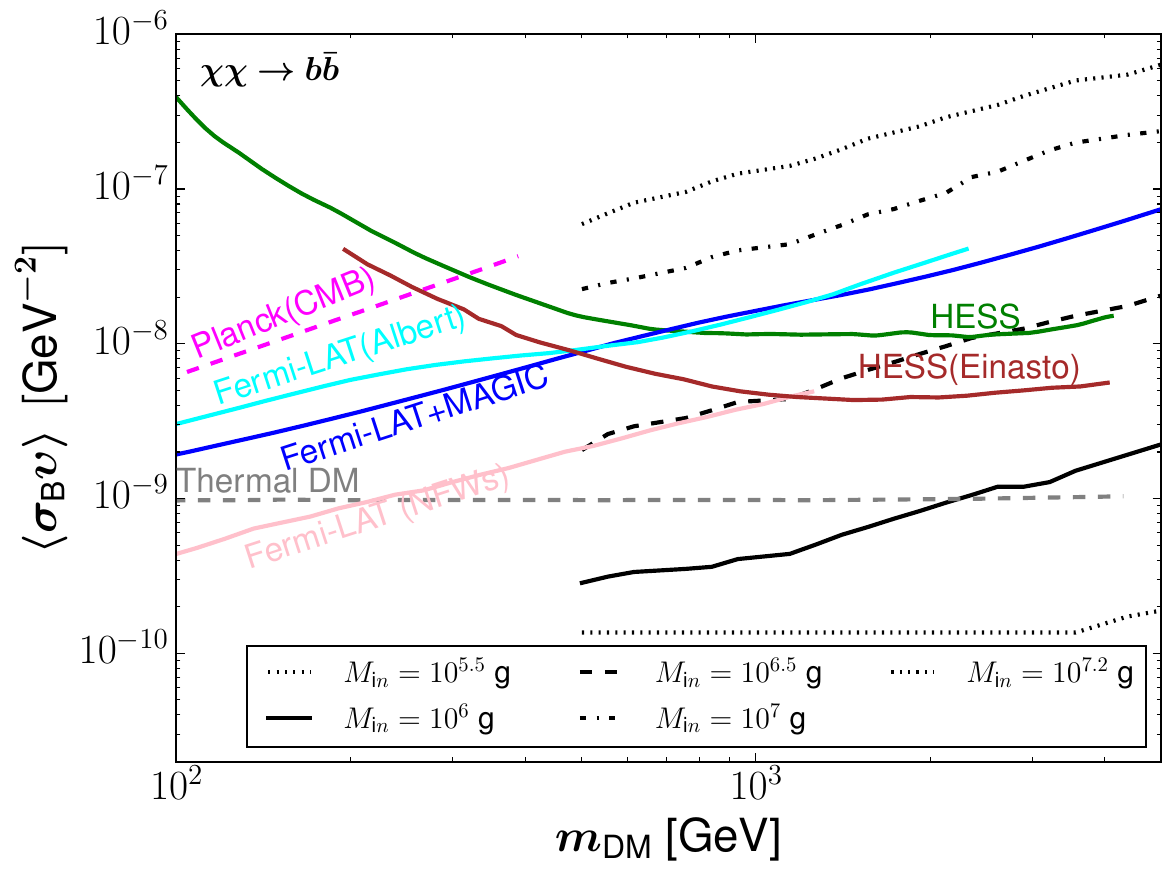}
\caption{\label{fig:cross section constraints} B-number violating annihilation cross section for producing observed baryon asymmetry with  DM masses for given  PBH masses. We show the  observational constraints from CMB analysis (magenta line)~\cite{Planck:2018vyg} and gamma ray measurements such as dwarf galaxies analysis from Fermi-LAT (cyan line)~\cite{Fermi-LAT:2016uux}, combined analysis from Fermi-LAT and Magic (blue line)~\cite{MAGIC:2016xys}, FERMI-LAT observations of the Inner Galaxy when DM distribution is assumed the compressed NFW profile (NFWs) (pink line)~\cite{German_2013} and H.E.S.S. observations of the galactic center (brown line)~\cite{HESS:2022ygk} corresponding to the assumption of Einasto profile of the DM distribution. The future observations are shown for the CTA collaboration~\cite{Carr:2016ct} due to the assumed NFW (dashed red line) and Einasto profiles (dashed orange line).
}
\end{figure}

In Fig.~\ref{fig:cross section constraints}, we show the B-number violating annihilation cross section $\langle \sigma_B v\rangle$ for the production of the observed baryon asymmetry with DM mass for given PBH masses $M_{\rm in}=10^{5.5}, 10^6, 10^{6.2}, 10^{6.5}, 10^7$, and $10^{7.2}$g, each represented by distinct lines as indicated in the labels. We compare them with the observational constraints for DMs annihilating into $b\bar{b}$ quarks from CMB analysis (magenta line)~\cite{Planck:2018vyg} and gamma ray measurements such as dwarf galaxies analysis from Fermi-LAT (cyan line)~\cite{Fermi-LAT:2016uux}, combined analysis from Fermi-LAT and Magic (blue line)~\cite{MAGIC:2016xys}, FERMI-LAT observations of the Inner Galaxy when DM distribution is assumed the compressed NFW profile (NFWs) (pink line)~\cite{German_2013} and H.E.S.S. observations of the galactic center (brown line)~\cite{HESS:2022ygk} corresponding to the assumption of Einasto profile of the DM distribution. Additionally, the future observations are shown for Cherenkov Telescope Array (CTA) collaboration~\cite{Carr:2016ct} taking into account for the assumed NFW (dashed red line) and Einasto profiles (dashed orange line). Note that here to accommodate the observational constraints on the annihilation cross section from the DM to our model, we consider the difference of the energy of annihilation products and number of quarks produced. In our case, only one quark is produced from DM annihilation with energy around $m_{\rm DM}- m_\psi^2/(4m_{\rm DM})$,
while usual constraints are put for the production of two quarks with energy of DM mass.

\section{Discussion}\label{con}
We have investigated a scenario where non-thermal WIMP DM from PBH evaporation can resolve both the relic abundance of DM and the observed baryon asymmetry at the same time. 
The PBH evaporation after thermal freeze-out of WIMP can produce the non-thermal WIMP which re-annihilate to explain correct relic density of DM. 
During the reannihilation, B and CP-violating DM annihilation can generate baryon asymmetry too. 
For this mechanism, larger annihilation cross section is needed compared to thermal freeze-out WIMP.
We find that the primordial black hole with mass less than $10^7 {\rm g}$ is a 
 good candidate as source of TeV dark matter with the total annihilation cross section $\left\langle\sigma_a \upsilon\right\rangle \lesssim 10^{-7} \ {\rm GeV}^{-2}$, and
 the B-violating cross section including one quark $\left\langle\sigma_B \upsilon\right\rangle \lesssim 2\times 10^{-9} \ {\rm GeV}^{-2}$. This upper bound comes from the gamma-ray search produced by DM annihilation in our galaxy. This indirect detection of the gamma-rays or cosmic rays also provide good methods to probe the other parameter spaces in our model in the near future.

\acknowledgments
The authors were supported by the National Research Foundation of Korea (NRF) grant funded by the Korea government (MEST) NRF-2022R1A2C1005050(KYC, EL) and NRF-2019R1A2C3005009 (JK).



\bibliographystyle{utphys}
\bibliography{cas-refs}
\end{document}